\documentclass[12pt]{article}
\usepackage[utf8]{inputenc}
\usepackage{jheppub}

\usepackage{parskip}
\usepackage{mathtools}
\usepackage{slashed}
\usepackage{braket}
\usepackage{enumitem}
\usepackage{xcolor}
\usepackage{colortbl}
\usepackage{adjustbox}
\usepackage{pifont}
\usepackage{float}
\usepackage[bottom]{footmisc}

\usepackage{tikz}
\usepackage{tikz-cd}
\usetikzlibrary{shapes.geometric}
\usetikzlibrary{scopes}
\usetikzlibrary{positioning}
\usetikzlibrary{calc}
\tikzstyle{block} = [draw=black, thick, text width=2cm, minimum height=1cm, align=center, fill = gray!10]

\numberwithin{equation}{section}

\newcommand{\bZ}{\mathbb{Z}}
\newcommand{\bS}{\mathbb{S}}
\newcommand{\bC}{\mathbb{C}}
\newcommand{\so}{\mathfrak{so}}
\newcommand{\Arf}{\mathrm{Arf}}
\newcommand{\PD}{\mathrm{PD}}
\newcommand{\NS}{\mathrm{NS}}
\newcommand{\R}{\mathrm{R}}
\newcommand{\cT}{\mathcal{T}}
\newcommand{\cI}{\mathcal{I}}
\newcommand{\cS}{\mathcal{S}}
\newcommand{\cZ}{\mathcal{Z}}
\newcommand{\cN}{\mathcal{N}}
\newcommand{\cH}{\mathcal{H}}
\newcommand{\Spin}{\mathrm{Spin}}
\newcommand{\SO}{\mathrm{SO}}

\newcommand{\pt}{\mathrm{pt}}
\newcommand{\CS}{\mathrm{CS}}
\newcommand{\grav}{\mathrm{grav}}
\newcommand{\CSgrav}{\CS_\grav}

\newcommand{\cola}{e5f6ff}
\newcommand{\colb}{e7ffcd}
\newcommand{\colc}{ffecf9}
\newcommand{\cold}{fff9d9}
\newcommand{\setbackground}[2]{\text{\colorbox[HTML]{#1}{$#2$}}}
\newcommand{\sect}[1]{\mathsf{#1}}
\newcommand{\secta}{\sect{A}}
\newcommand{\sectb}{\sect{B}}
\newcommand{\sectc}{\sect{C}}
\newcommand{\sectd}{\sect{D}}
\newcommand{\backa}{\setbackground{\cola}{\secta}}
\newcommand{\backb}{\setbackground{\colb}{\sectb}}
\newcommand{\backc}{\setbackground{\colc}{\sectc}}
\newcommand{\backd}{\setbackground{\cold}{\sectd}}
\newcommand{\cella}{\cellcolor[HTML]{\cola}\secta}
\newcommand{\cellb}{\cellcolor[HTML]{\colb}\sectb}
\newcommand{\cellc}{\cellcolor[HTML]{\colc}\sectc}
\newcommand{\celld}{\cellcolor[HTML]{\cold}\sectd}

\newcommand{\spinarg}[3]{\begin{tikzpicture}[scale=0.4, baseline=0] \draw (0,0) rectangle (1,1); \node[anchor=north, inner sep=0mm] at (0.5,-0.2) {\scriptsize{#2}}; \node[anchor=east, inner sep=0mm] at (-0.1,0.5) {\scriptsize{#1}}; #3 \end{tikzpicture}}
\newcommand{\spin}[2]{\spinarg{#1}{#2}{}}
\newcommand{\spinx}[2]{\spinarg{#1}{#2}{\draw (0,0.5) -- (1,0.5);}}
\newcommand{\spiny}[2]{\spinarg{#1}{#2}{\draw (0.5,0) -- (0.5,1);}}
\newcommand{\spinxy}[2]{\spinarg{#1}{#2}{\draw (0,0.5) -- (1,0.5); \draw (0.5,0) -- (0.5,1);}}
\newcommand{\greent}{green!65!black}
\definecolor{dgreen}{rgb}{0, 0.7, 0}

\title{Backfiring Bosonisation}

\author[a]{Philip Boyle Smith}
\author[a,b,c]{and Yunqin Zheng}

\affiliation[a]{
Kavli Institute for the Physics and Mathematics of the Universe (WPI), \\
University of Tokyo, Kashiwa, Chiba 277-8583, Japan
}

\affiliation[b]{
Institute for Solid State Physics, University of Tokyo, Kashiwa, Chiba 277-8583, Japan
}

\affiliation[c]{
C. N. Yang Institute for Theoretical Physics, Stony Brook University, Stony Brook, NY 11794, USA
}

\emailAdd{philip.boyle.smith@ipmu.jp}
\emailAdd{yunqin.zheng@stonybrook.edu}

\abstract{
For a fermionic quantum field theory in $d=1+1$ dimensions, there is a subtle difference between summing over spin structures and gauging $(-1)^F$.
If the gravitational anomaly vanishes mod 16, then both operations are equivalent and yield a bosonic theory. But if the gravitational anomaly only vanishes mod 8, then only gauging $(-1)^F$ is allowed, and the result is a fermionic theory.
Our goal is to understand in detail how this happens, despite the fact $(-1)^F$ is defined in terms of shifting the spin structure, which would naïvely suggest that both operations are equivalent.
We do this from three perspectives: an abstract view in terms of anomalies, explicit CFT calculations, and a Symmetry TFT perspective.
To conclude, we illustrate our results using the heterotic string and the famous self-triality of 8 Majorana-Weyl fermions.
}

\begin{document}

\setcounter{tocdepth}{2}
\maketitle

\newpage

\section{Introduction}\label{sec:int}

A fermionic quantum field theory, by definition, depends upon a background spin structure. There are a number of interesting operations one can perform by manipulating this spin structure, allowing one to generate new theories from old. These give rise to various 2d dualities \cite{Coleman:1974bu, Witten:1983ar, ytlecture, web2d, ttconfmfd, orbifoldgroupoids}, which play a role in the worldsheet of the superstring \cite{Seiberg:1986by, tscworldsheet, Kaidi:2019pzj, 8maj}, the study of generalised symmetries and topological phases \cite{ttconfmfd, Kaidi:2021xfk, Thorngren:2021yso, Yao:2019bub, Fukusumi:2021zme, Yao:2020dqx, Lin:2019hks, Hsieh:2020uwb, Kulp:2020iet, Ji:2019ugf, Thorngren:2018bhj, Debray:2023iwf}, and symmetric mass generation \cite{8maj, Fidkowski:2009dba, BenTov:2014eea, Tong:2021phe, Wang:2022ucy, Zeng:2022grc, PhysRevX.8.011026}.

Two fundamental, and apparently identical such operations are summing over spin structures and gauging fermion parity $(-1)^F$. The reason why one might think these operations are identical is because the gauge field for $(-1)^F$ simply acts by shifting the spin structure, and so summing over all gauge fields is equivalent to summing over all spin structures. Since the resulting theory is manifestly bosonic, this is called \emph{bosonisation}.\footnote{In the literature \cite{Coleman:1974bu, Witten:1983ar}, the term bosonisation was often used as a duality/equivalence between a special pair of theories, e.g.\ between the Thirring and Sine-Gordon models. In this work, we refer to bosonisation as a map that can be applied to a generic fermionic theory. The difference and relations between these two perspectives has been discussed recently in \cite{web2d, Ji:2019ugf, Hsieh:2020uwb}.}

The purpose of this paper is to understand a number of subtleties in the above story that arise when the theory has a gravitational anomaly---in particular a gravitational anomaly of 8 mod 16. For such a theory, $(-1)^F$ is still anomaly-free, so can be gauged. But upon gauging it, one finds that the result is another fermionic theory---that is, the attempt to bosonise has backfired.

The only way out of the above paradox is if spin structure summation and gauging fermion parity are in fact different operations.  And indeed, one quick way to see this is that their anomalies are different: as we will review, the anomaly of the spin structure is mod-16 valued, while the anomaly of $(-1)^F$ is mod-8 valued, so they cannot be the same. Thus the name ``bosonisation'' should be reserved for spin structure summation, which always produces a bosonic theory. Gauging fermion parity is equivalent to it when both can be defined, namely when the anomaly is 0 mod 16, but gauging $(-1)^F$ is also possible when the anomaly is 8 mod 16, where it is a different kind of operation that produces a fermionic theory, for which a better title might be \emph{refermionisation}.

That doesn't get us entirely off the hook, however, as several puzzles remain to be explained. For a start, the equivalence of spin structure summation to gauging $(-1)^F$ in the anomaly-free case follows from an extremely simple calculation (which we review momentarily), and it is not immediately obvious how this is modified by the anomaly, and only for particular values. Moreover in the 8 mod 16 case, there is no obvious formula which can turn an old fermionic partition function into a new one, aside from stacking it with the Arf invariant, but this is not the right answer. The fact that $(-1)^F$ is defined in terms of shifting the spin structure, yet has a different anomaly to it, also requires care to reconcile.

\subsection{Summary of Results}

Our goal is to revisit the operations of spin structure summation and gauging fermion parity in gravitationally-anomalous theories, while giving a careful resolution of the issues we have just described from a modern perspective.

We begin by considering the effect of the gravitational anomaly. Due to the SPT bulk, the usual definition of $(-1)^F$ in terms of shifting the spin structure is insufficient, and we show how this definition must be amended. From this follows the non-equivalence of spin structure summation and gauging $(-1)^F$, the difference in their anomalies, and the possibility for gauging $(-1)^F$ to produce a fermionic theory. We also learn that there is no canonical way to gauge $(-1)^F$ in a theory with gravitational anomaly 8 mod 16, but instead two inequivalent ways.

To complement the preceding formal discussion, we illustrate it with explicit CFT calculations on the torus. Here the gravitational anomaly manifests as anomalous phases under modular $\cS$ and $\cT$ transformations in a standard way, allowing the previous claims to be realised very concretely. We use this perspective to derive the duality webs obeyed by gauging $(-1)^F$, which follows a triality-like structure when the anomaly is 8 mod 16.

A more modern way to understand duality webs comes from Symmetry TFT, in this case the $\Spin(n)_{-1}$ Chern-Simons theory. We discuss the invertible condensation defects in this theory, and the topological interfaces to SPT phases obtained from condensing bosonic or fermionic lines. We interpret the operations of bosonisation, refermionisation, and stacking an Arf invariant as the fusion of various condensation defects with the topological interfaces.

The above issues are crucial to the correct formulation of the self-triality of 8 chiral Majorana-Weyl fermions. As an application, we note that this triality actually follows from the universal duality web obeyed by refermionisation, in combination with the uniqueness of a chiral fermionic CFT at $c = 4$. We also show how this web is realised on the classification of chiral fermionic CFTs of low central charge $c \leq 12$.

\subsection{The Plan of the Paper}

The paper is structured as follows. In Section~\ref{sec:review}, we review some motivational facts about bosonisation and provide more background. Sections~\ref{sec:anomalies}-\ref{sec:symtft} contain our main results, with Section~\ref{sec:anomalies} dedicated to the effect of gravitational anomalies on spin structure summation and gauging $(-1)^F$, Section~\ref{sec:cft} to explicit CFT calculations on the torus, and Section~\ref{sec:symtft} to the Symmetry TFT analysis. In Section~\ref{sec:applications} we conclude with some applications to chiral triality and fermionic CFTs of low central charge.

\section{Review of Bosonisation}\label{sec:review}

This paper is concerned with spin structure summation and gauging fermion parity, and the subtle difference between them in the presence of a gravitational anomaly. It will therefore be useful to begin with a review of these operations in gravitationally-non-anomalous theories, where they are the same.

We start with a fermionic quantum field theory $T_F$ in $d=2$ dimensions with no gravitational anomaly. Such a theory is described by a partition function $\cZ_F[\Sigma, g, \rho]$ that depends upon an oriented surface $\Sigma$, a Riemannian metric $g$, and a spin structure $\rho$. For the most part, we can assume $\Sigma$ and $g$ are held fixed, in which case we can abbreviate the partition function to simply $\cZ_F[\rho]$.

The operation of spin structure summation turns $T_F$ into a bosonic theory $T_B$ with $\bZ_2$ symmetry. At the level of partition functions, the transformation is
\begin{equation}\label{eq:spinsumdef}
    \cZ_B[A] = \frac{1}{\sqrt{|H^1(\Sigma; \bZ_2)|}} \sum_\rho (-1)^{\Arf[\rho + A] + \Arf[\rho]} \cZ_F[\rho]
\end{equation}
where $A$ is the background field for the $\bZ_2$ symmetry. The most important feature of \eqref{eq:spinsumdef} is the Arf invariant of a spin structure, $\Arf[\rho]$, a mod-2 valued quantity given by the mod-2 index of the Dirac operator. It appears via $(-1)^{\Arf[\rho]}$, which is the partition function of an SPT phase we will refer to as the Arf theory, though it is better known to condensed matter theorists as the topological phase of the Kitaev chain. For a nice review of the properties and uses of Arf, see \cite{web2d}.

Formula \eqref{eq:spinsumdef} also makes use of the ability to add a $\bZ_2$ gauge field $A$ to a spin structure $\rho$ to produce another spin structure $\rho + A$. In fact, if we pick a reference spin structure $\rho_0$, then any $\rho$ can be written uniquely as $\rho = \rho_0 + A$ for some $A$. Mathematicians would say that the space of spin structures forms an affine space modelled on the space of $\bZ_2$ gauge fields $H^1(\Sigma; \bZ_2)$, meaning these spaces are isomorphic but only non-canonically (due to the need to choose $\rho_0$). Thus the functions $Z_B[A]$ and $Z_F[\rho]$ are superficially similar, yet different kinds of objects.

Although we have phrased the operation as a transformation of partition functions, it can also be lifted to a transformation of the full quantum field theories. To do this, we write it in the form
\[
    T_B = (\Arf \times \Arf \times T_F) \; / \; \text{Spin Structure}
\]
The $\bZ_2$ gauge field of $T_B$ shifts the spin structure of the first $\Arf$. However, for our purposes, nothing will be lost by focusing only on partition functions.

Starting from the fermionic theory $T_F$, one can twist the bosonisation by further stacking an Arf before summing over the spin structure. The resulting bosonic theory $T_{B'}$ is related to $T_B$ via the Kramers-Wannier transformation $T_{B'} = T_B / \bZ_2$ \cite{ytlecture, Karch:2019lnn, Hsieh:2020uwb, Ji:2019ugf}. In this work, we will mostly focus on $T_B$.

We also have the operation of gauging fermion parity. This proceeds by taking $\cZ_F[\rho]$ and first coupling it to a background field for $(-1)^F$, yielding $\cZ_F[\rho + A]$. We then allow $A$ to fluctuate, indicating this by renaming it $a$. We can also include a coupling to a background field $A$ for the quantum, emergent $\bZ_2$ symmetry \cite{Vafa:1989ih}. The result is
\begin{equation}\label{eq:gaugingdef}
    \cZ_{F'}[\rho, A] = \frac{1}{\sqrt{|H^1(\Sigma; \bZ_2)|}} \sum_a (-1)^{\int_\Sigma a \smile A} \cZ_F[\rho + a]
\end{equation}
Note that $T_{F'} = T_F / (-1)^F$ is a fermionic theory with a specified choice of $\bZ_2$ symmetry, and so initially appears very different to $T_B$. Nonetheless \eqref{eq:gaugingdef} also includes a sum over spin structures, since as $a$ is summed over, $\rho + a$ takes on every spin structure exactly once. We can make this similarity manifest by making use of the identity (see Appendix~\ref{app:arf})
\begin{equation}\label{eq:arfidentity}
    \int_\Sigma a \smile A = \Arf[\rho + A] + \Arf[\rho] + \Arf[\rho + a + A] + \Arf[\rho + a]
\end{equation}
Substituting \eqref{eq:arfidentity} into \eqref{eq:gaugingdef}, we find that $T_{F'}$ is related to $T_B$ as
\begin{equation}\label{eq:FpB}
    \cZ_{F'}[\rho, A] = (-1)^{\Arf[\rho + A] + \Arf[\rho]} \cZ_B[A]
\end{equation}
The factor sitting in front of $\cZ_B[A]$ is a mere topological counterterm, and moreover one which is trivial when $A = 0$. We are therefore free to drop it as part of the freedom in the definition of gauging. We learn that despite its fermionic appearance, up to a counterterm $T_{F'}$ is actually a bosonic theory in disguise, and in fact the same as $T_B$. It's for this reason that gauging fermion parity, at least in the gravitationally anomaly-free case, is said to be equivalent to spin structure summation.

The relation between summing over spin structures and gauging fermion parity has also been discussed in $d=3$ dimensions \cite{lorentzfrac}. In contrast to the situation in 2d, the gravitational anomaly in 3d is always trivial, so the subtleties we are interested in do not arise. Indeed, in \cite{lorentzfrac} the authors found that the two operations only differ by stacking a fermionic SPT $\SO(r)_1$. Hence by absorbing this invertible theory into the definition of gauging fermion parity, the two operations are the same in disguise.

\subsection{An Example: The Ising Model}

The archetypical example of bosonisation is the relation between the Ising model and the Majorana fermion \cite{Ji:2019ugf, Karch:2019lnn, Lin:2019hks}. For us, this example will also be useful to highlight how bosonisation acts on the torus, a topic we return to in Section~\ref{sec:cft}.

We start by taking $T_F$ to be a single Majorana fermion, a CFT of central charge $c = \bar{c} = \frac{1}{2}$. The gravitational anomaly is $n = 2(c - \bar{c}) = 0$, so the above discussion applies. We place the theory on $\Sigma = T^2$, and separate the states into four types depending on whether the spin structure of the spatial circle is NS or R, and also their fermion parity. These states can be arranged into a table
\[
    \mathllap{T_F \, :} \qquad
    \begin{array}{r|cc}
            & + & - \\
        \hline
        \NS & \cella & \cellb \\
        \R  & \cellc & \celld
    \end{array}
\]
where $\secta, \sectb, \sectc, \sectd$ denote the various twisted sectors. For the Majorana fermion, they are equal to
\begin{alignat*}{1}
    \backa &= |\chi_0|^2 + |\chi_{1/2}|^2 \\
    \backb &= \chi_0 \overline{\chi}_{1/2} + \chi_{1/2} \overline{\chi}_0 \\
    \backc = \backd &= |\chi_{1/16}|^2
\end{alignat*}
in terms of the Virasoro characters at $c = \frac{1}{2}$. If we trace through the definition \eqref{eq:spinsumdef}, we find that the effect of bosonisation is to permute the twisted sectors, yielding
\[
    \mathllap{T_B \, :} \qquad
    \begin{array}{r|cc}
                         & + & - \\
        \hline
        \text{untwisted} & \cella & \cellc \\
        \text{twisted}   & \celld & \cellb
    \end{array}
\]
This is the Ising CFT. Indeed, all the states in the untwisted sector, i.e.\ $\secta$ and $\sectc$, have integer spins, and hence we have a bosonic theory. As such, the interpretation of the table has now shifted to match, with the rows labelled by the holonomy of the $\bZ_2$ background around the spatial circle and the columns by the $\bZ_2$ charge. Thus although the table superficially resembles that of $T_F$, it is still a different kind of object, tallying with our earlier remarks on the partition functions $\cZ_F[\rho]$ and $\cZ_B[A]$.

\subsection{An Anti-Example: 8 Majorana-Weyl Fermions}

Everything so far carries through unchanged for a theory with gravitational anomaly $n = 0$ mod 16 \cite{Seiberg:1986by}. But it is a beloved fact going back to Fidkowski and Kitaev \cite{Fidkowski:2009dba, Ryu:2012he} that the anomaly of fermion parity is a mod-8 effect, not a mod-16 effect. Indeed, treating the fermion parity symmetry $(-1)^F$ as a generic $\bZ_2$ symmetry, the anomaly of a $\bZ_2$ symmetry in a 2d spin theory is classified by the reduced Anderson dual of the spin bordism group $(\widetilde{I_\bZ\Omega^\Spin})^4(B\bZ_2) = \bZ_8$ \cite{Yamashita:2021cao,Lee:2020ojw, Freed:2016rqq, Kapustin:2014dxa, Wan:2018bns}. Therefore it should also be possible to gauge fermion parity in theories with $n = 8$ mod 16, and one would naïvely expect the same conclusions to apply.

To see if this is the case, let us take $T_F$ to be the theory of 8 chiral Majorana-Weyl fermions, a CFT with $(c, \bar{c}) = (4, 0)$. The gravitational anomaly is $n = 2(c - \bar{c}) = 8$, which places us in the interesting case $n = 8$ mod 16. Re-running the arguments of the previous section, the table of twisted sectors of $T_F$ is
\[
    \mathllap{T_F \, :} \qquad
    \begin{array}{r|cc}
            & + & - \\
        \hline
        \NS & \cellcolor[HTML]{\cola} \chi_1 & \cellcolor[HTML]{\colb} \chi_V \\
        \R  & \cellcolor[HTML]{\colc} \chi_S & \cellcolor[HTML]{\cold} \chi_C
    \end{array}
\]
where the entries are the characters of $\widehat{\so}(8)_1$. The NS sector contains states with half-integer spins from $\chi_V$, reflecting the theory being fermionic. It follows that the gauged theory $T_B$ is
\begin{equation}\label{eq:TB}
    \mathllap{T_B ? \, :} \qquad
    \begin{array}{r|cc}
                         & + & - \\
        \hline
        \text{untwisted} & \cellcolor[HTML]{\cola} \chi_1 & \cellcolor[HTML]{\colc} \chi_S \\
        \text{twisted}   & \cellcolor[HTML]{\cold} \chi_C & \cellcolor[HTML]{\colb} \chi_V
    \end{array}
\end{equation}
But by $\so(8)$ triality, this is none other than $T_F$ up to a reparametrisation of the global $\Spin(8)$ symmetry, since the cyclic permutation of $(\chi_V, \chi_C, \chi_S)$ can be undone by an outer automorphism of $\so(8)$. We can also see the fermionic nature of $T_B$ from the untwisted sector, which contains half-integer spins from $\chi_S$. This illustrates the ``backfiring bosonisation'' phenomenon. Clearly, the above table wants to be
\begin{equation}\label{eq:TFp}
    \mathllap{T_{\widetilde{F}} \, :} \qquad
    \begin{array}{r|cc}
            & + & - \\
        \hline
        \NS & \cellcolor[HTML]{\cola} \chi_1 & \cellcolor[HTML]{\colc} \chi_S \\
        \R  & \cellcolor[HTML]{\cold} \chi_C & \cellcolor[HTML]{\colb} \chi_V
    \end{array}
\end{equation}
where $T_{\widetilde{F}}$ is a \emph{new} fermionic theory. But that is not what \eqref{eq:gaugingdef} hands us. A partition function $\cZ_B[A]$ cannot be converted into a partition function $\cZ_{\widetilde{F}}[\rho]$ without a choice of reference spin structure. For example, on the torus, one can choose the $\rho_{\NS,\NS}$ reference spin structure, allowing one to write down the relation
\[
    \cZ_{\widetilde{F}}[\rho] = \cZ_B[A = \rho - \rho_{\NS,\NS}]
\]
However, there is nothing canonical about the choice $\rho_{\NS,\NS}$, and furthermore the problem is compounded at higher genus, where there is no obvious way to make such a choice. One of our goals in Section~\ref{sec:anomalies} is to understand the transformation from $T_F$ to $T_{\widetilde{F}}$ directly, without visiting a bosonic-but-actually-fermionic theory $T_B$ in between, both at arbitrary genus and without the need to pick a reference spin structure.

\subsection{General Chiral Theories}

To finish this section, we note a general pattern among our examples. In a generic theory $T_F$ with gravitational anomaly $n \in \bZ$,\footnote{This characterisation of the anomaly requires unitarity, which we assume throughout.} the topological spins of the various twisted sectors are
\[
    \backa = 0 \qquad
    \backb = \frac{1}{2} \qquad
    \backc = \backd = \frac{n}{16}
\]
We see that $T_B$ contains operators of spins 0 and $\frac{n}{16}$ in its untwisted sector. Thus $T_B$ is bosonic for $n = 0$ mod 16 and fermionic for $n = 8$ mod 16. Note that the topological spins above are those of the Chern-Simons theory $\Spin(n)_{-1}$; this is because $\Spin(n)_{-1}$ is the Symmetry TFT for generalised gauging operations (or topological manipulations \cite{orbifoldgroupoids}) involving the spin structure. We return to this topic in Section~\ref{sec:symtft}.

\section{Anomalies and Spin Structures}\label{sec:anomalies}

In this section we analyse the effect of the gravitational anomaly on the equivalence between gauging $(-1)^F$ and summing over spin structures.

Before getting into the weeds, here is a simple illustration of why the gravitational anomaly must play an essential role. An anomalous fermionic theory $T_F$ does not, in fact, have a phase-unambiguous partition function merely depending on $(\Sigma, g, \rho)$ as originally claimed in Section~\ref{sec:review}. In order to fix the phase of the partition function, one must also pick a manifold of one higher dimension $M_3$ with boundary $\Sigma$ over which these structures extend. For example, if we take $(\Sigma, \rho)$ to be $T^2$ with the NS-R spin structure, then we can write $(\Sigma, \rho)$ as a boundary in the following way:
\[
    (S^1)_\NS \times (S^1)_\R = \partial \big[ \, D^2 \times (S^1)_\R \, \big]
\]
Now suppose we keep the 2-manifold $\Sigma$ the same, but change its spin structure to R-NS. Then it is not enough to simply change the spin structure on $M_3$ to match; we must also change the topology of $M_3$. In this case a suitable 3-manifold is
\[
    (S^1)_\R \times (S^1)_\NS = \partial \big[ \, (S^1)_\R \times D^2 \, \big]
\]
This shows that when dealing with fermion parity and spin structures, it is not enough to sweep the bulk manifold $M_3$ under the rug, but instead it must be embraced.

As a further complication, some choices of $(\Sigma, \rho)$ do not even admit a choice of $M_3$. The simplest example is $T^2$ with the R-R spin structure, which is not the boundary of any spin 3-manifold, as detected by its nontrivial Arf invariant. To deal with this issue, we must use the formulation of anomalous theories involving bulk SPT phases \cite{Freed:2012bs}. We review this for the relevant gravitational anomalies in the next subsection.

\subsection{Gravitational Anomalies}\label{sec:grav}

According to the general correspondence between anomalies and SPT phases \cite{Freed:2012bs, wittenphases}, the gravitational anomaly of a 2d fermionic theory $T_F$ is related to a fermionic SPT phase in one dimension higher. Such fermionic 3d SPT phases are classified by $(I_\bZ\Omega^\Spin)^4(\pt) = \bZ$, where the SPT associated with $n \in \bZ$ is the $\SO(n)_{-1}$ Chern-Simons theory. This theory has action given by a gravitational Chern Simons term with level $n$, i.e.
\begin{equation}\label{eq:fer_SPT}
    \cZ_{\SO(n)_{-1}}[M_3, g_3, \rho_3] = \exp \! \left( 2 \pi i n \int_{M_3} \CSgrav \right)
\end{equation}
Here $M_3$ is an oriented 3-manifold (all such manifolds are spin), $g_3$ is a metric, and $\rho_3$ is a choice of spin structure.

The right hand side of \eqref{eq:fer_SPT} doesn't obviously depend on $\rho_3$. To see this hidden dependence, we can equivalently rewrite it as
\begin{equation}\label{eq:fer_SPT_Ahat}
    \cZ_{\SO(n)_{-1}}[M_3, g_3, \rho_3] = \exp \! \left( 2 \pi i n \int_{M_4} \frac{1}{2} \hat{A}(R) \right)
\end{equation}
To use this definition one has to pick a 4-manifold $(M_4, g_4, \rho_4)$ with boundary $(M_3, g_3, \rho_3)$ over which all the structures extend. This is always possible since $\Omega_3^\Spin(\pt) = 0$. By a similar argument to last time, changing $\rho_3$ then requires us to change the topology of $M_4$, which affects the value of the integral.

We briefly comment on why the result is independent of the choice of $(M_4, g_4, \rho_4)$. The index theorem states that $\int_{M_4} \hat{A}(R) = \mathrm{index}(\slashed{D})$ for a closed spin manifold $M_4$. Because the Dirac operator in $d=4$ dimensions has a Kramers-doubled spectrum via $\psi \rightarrow \gamma^0 \gamma^2 \psi^*$, the index is even, and so $\int_{M_4} \frac{1}{2} \hat{A}(R)$ is an integer. By the standard gluing argument \cite{wittenphases}, it follows that \eqref{eq:fer_SPT_Ahat} is independent of the choice of extension.

The theory $\SO(n)_{-1}$ is always fermionic. But it is sometimes a bosonic theory in disguise. For a closed oriented 4-manifold $M_4$, there are the relations
\[
    \hat{A}(R) = -\frac{1}{24} p_1 \qquad L_1 = \frac{1}{3} p_1 \qquad \int_{M_4} L_1 = \text{signature}(M_4)
\]
These imply $16 \int_{M_4} \frac{1}{2} \hat{A}(R)$ is an integer on any closed oriented 4-manifold without requiring it to be spin. It follows that when $16|n$, there is a bosonic SPT with the same action as $\SO(n)_{-1}$. It is
\begin{equation}\label{eq:bos_SPT}
    \cZ_{\Spin(n)_{-1} / \bZ_2^S}[M_3, g_3] = \exp \! \left( 2 \pi i  n \int_{M_3} \CSgrav \right) \quad \text{for } 16|n
\end{equation}
The two SPTs are related via $\SO(n)_{-1} = \Spin(n)_{-1} / \bZ_2^S \times \SO(0)_1$, where $\SO(0)_1$ is the almost trivial theory with a single transparent fermionic line; this is what we mean by saying that $\SO(n)_{-1}$ is $\Spin(n)_{-1} / \bZ_2^S$ in disguise. The notation $/\bZ_2^S$ means gauging the $\bZ_2$ one-form symmetry generated by the bosonic line in the spinor representation of $\Spin(n)$, which is only bosonic if $n$ is a multiple of 16. Note that $\SO(n)_{-1}$ can also be expressed as $\Spin(n)_{-1} / \bZ_2^V$, but the $\bZ_2$ is generated by the fermionic line in the vector representation of $\Spin(n)$. Condensing a fermionic line of $\Spin(n)_{-1}$ leads to a spin theory $\SO(n)_{-1}$ \cite{Seiberg:2016rsg, Bhardwaj:2016clt}.

For us, the relevance of \eqref{eq:bos_SPT} is that it is the most general gravitational anomaly of a bosonic theory $T_B$. We can verify this using the duality $\Spin(16)_{-1}/\bZ_2^S \leftrightarrow (E_8)_{-1}$, the latter of which is well known as the minimal bosonic gravitational anomaly \cite{Kaidi:2021gbs, KitaevE8}.

The case of $n = 8$ mod 16 exhibits a particular subtlety that is worth highlighting before we go on. The most general anomaly action for a $\bZ_2$ symmetry of a 2d spin theory is
\[
    \cZ_\text{anom}[M_3, \rho_3, A_3] = \frac{\cZ_{\SO(n)_{-1}}[M_3, g_3, \rho_3 + A_3]}{\cZ_{\SO(n)_{-1}}[M_3, g_3, \rho_3]}
\]
where $A_3$ is the background field for the $\bZ_2$ symmetry and $n$ is the anomaly. When $n = 0$ mod 8, this anomaly is trivial, and the above action is equal to 1. This implies $\cZ_{\SO(n)_{-1}}[M_3, g_3, \rho_3]$ is independent of spin structure for $n = 0$ mod 8. However, despite the \emph{action} having the appearance of being bosonic, there is no bosonic \emph{theory} with this action unless additionally $n = 0$ mod 16. This fact was also recently noted in \cite{Barkeshli:2023bta}.

In summary, a fermionic theory has gravitational anomaly classified by a single integer $n$, which implies an SPT bulk phase of $\SO(n)_{-1}$ with action $n \CSgrav$. A bosonic theory also has gravitational anomaly classified by a single integer $n_B$, which implies an SPT bulk phase of $\Spin(16n_B)_{-1} / \bZ_2^S$ with action $16 n_B \CSgrav$.

\subsection{Coupling to Fermion Parity}\label{sec:coupling}

We are now in a position to deal with the main issues outlined in the introduction.

We start with a fermionic theory $T_F$ of gravitational anomaly $n$. The anomaly means the partition function is an element of the SPT Hilbert space\footnote{Although $\cZ_F[\Sigma, \rho]$ is a state, we have opted to avoid the traditional ket notation $\ket{\cZ_F[\Sigma, \rho]}$ used for states, since for partition functions it is common to suppress it.}
\begin{equation}\label{eq:state}
    \cZ_F[\Sigma, \rho] \in \cH_{\SO(n)_{-1}}[\Sigma, \rho]
\end{equation}
where we suppress the metric dependence for simplicity. The right hand side is isomorphic to $\bC$, but non-canonically. This means the partition function suffers a phase ambiguity if we try to define it as a complex number. We can try to resolve this ambiguity by picking an extending 3-manifold $M_3$, and the path integral of $\SO(n)_{-1}$ will then prepare a state $\bra{M_3}$ whose inner product with $\cZ_F[\Sigma, \rho]$ is a genuine complex number. However, as alluded to earlier, a choice of $M_3$ does not always exist, and so we will prefer to leave the state $\cZ_F[\Sigma, \rho]$ as it is.

\subsubsection*{Summing over Spin Structures}

To warm up, let us review the well-known fact that spin structure summation requires $n = 0$ mod 16. Let $k$ be the smallest integer such that $nk$ is a multiple of 16, namely $k = \frac{16}{\gcd(16, n)}$. Then taking the $k$th tensor power of both sides of \eqref{eq:state},
\begin{equation}\label{eq:state_pow_k}
    \cZ_F[\Sigma, \rho]^k \in \cH_{\SO(nk)_{-1}}[\Sigma, \rho]
\end{equation}
where we have used the fact that $\SO(n)_{-1} \times \SO(m)_{-1} = \SO(n + m)_{-1}$. Because $nk$ is a multiple of 16, $\SO(nk)_{-1}$ is nothing more than the bosonic theory $\Spin(nk)_{-1}/\bZ_2^S$ regarded as a spin theory, so we can equally write
\begin{equation}\label{eq:state_pow_k_bos}
    \cZ_F[\Sigma, \rho]^k \in \cH_{\Spin(nk)_{-1}/\bZ_2^S}[\Sigma]
\end{equation}
Crucially the right hand side is now the Hilbert space of a bosonic theory, and so does not depend on $\rho$. Equation~\eqref{eq:state_pow_k_bos} shows that the more even $n$ is, the less anomalous $\cZ_F[\Sigma, \rho]$ is as a function of spin structure.

In order to be able to sum over spin structures as in \eqref{eq:spinsumdef}, we need to be able to form the sum
\[
    \sum_\rho \cZ_F[\Sigma, \rho]
\]
But each term lives in a different Hilbert space $\cH_{\SO(n)_{-1}}[\Sigma, \rho]$. It does not make sense to add such objects together. The only way is if the different Hilbert spaces are actually canonically isomorphic. To answer this we turn to equation~\eqref{eq:state_pow_k_bos}, which states that they are canonically isomorphic up to $k$th roots of unity. Thus only for $k = 1$, corresponding to $n$ a multiple of 16, does the sum exist, and in that case \eqref{eq:state_pow_k_bos} states that it lies in
\[
    \sum_\rho \cZ_F[\Sigma, \rho] \in \cH_{\Spin(n)_{-1}/\bZ_2^S}[\Sigma]
\]
We see that this is the correct space for a would-be bosonic partition function $\cZ_B[\Sigma]$ to live in, as expected.

\subsubsection*{The Definition of Fermion Parity}

Now we would like to couple the theory $T_F$ to a background for fermion parity. This requires us to define the partition function $\cZ_F[\Sigma, \rho, A]$ with a $(-1)^F$ background $A$, treated as an independent $\bZ_2$ symmetry. In the anomaly-free case \eqref{eq:gaugingdef}, this was as simple as defining
\begin{equation}\label{eq:couplingwrongdef}
    \cZ_F[\Sigma, \rho, A] \coloneqq \cZ_F[\Sigma, \rho + A]
\end{equation}
However \eqref{eq:couplingwrongdef} has a potential problem in anomalous theories. Since we ultimately wish to gauge $A$, we require \eqref{eq:couplingwrongdef} to be an \emph{anomaly-free} regularisation of $(-1)^F$, which in this context means that the Hilbert space it belongs to does not vary with $A$. That is, we require our definition to satisfy
\[
    \cZ_F[\Sigma, \rho, A] \in \cH_{\SO(n)_{-1}}[\Sigma, \rho]
\]
Note that the Hilbert space is however still allowed to vary with $\rho$, since the coupled theory still has the same gravitational anomaly. But the right-hand side of \eqref{eq:couplingwrongdef} does not have this property. Instead, it belongs to
\[
    \cZ_F[\Sigma, \rho + A] \in \cH_{\SO(n)_{-1}}[\Sigma, \rho + A]
\]
This Hilbert space is generically not equal to the previous one, being canonically isomorphic only for $n = 0$ mod 16. Thus the straightforward definition of fermion parity \eqref{eq:couplingwrongdef} only works for $n = 0$ mod 16.

This poses an interesting puzzle for the case $n = 8$ mod 16. Since $n = 0$ mod 8, $(-1)^F$ is anomaly-free by virtue of the fact that $(\widetilde{I_\bZ\Omega})^\Spin_4 = \bZ_8$, and so it should still be possible to define $\cZ_F[\Sigma, \rho, A]$. But if not by \eqref{eq:couplingwrongdef}, then how? The answer is that if we limit ourselves to writing geometrically valid equations that don't compare elements in different Hilbert spaces, then equation~\eqref{eq:state_pow_k_bos} says that the best we can do is
\begin{equation}\label{eq:couplingseconddef}
    \cZ_F[\Sigma, \rho, A]^2 \coloneqq \cZ_F[\Sigma, \rho + A]^2
\end{equation}
This is a strange definition. It leaves $\cZ_F[\Sigma, \rho, A]$ undetermined up to a sign for every possible value of its arguments. Fortunately, we can supplement \eqref{eq:couplingseconddef} with an additional requirement: that $\cZ_F[\Sigma, \rho, A]$ is compatible with cutting and pasting. This will generate relations among the various sign ambiguities, and it is not even obvious that a solution exists. However, courtesy of the fact $(-1)^F$ is anomaly-free, a solution is guaranteed.

We can now ask how unique our solution for $\cZ_F[\Sigma, \rho, A]$ is. The answer is that it is unique up to
\begin{equation}\label{eq:freedom}
    \cZ_F[\Sigma, \rho, A] \rightarrow (-1)^{n_1 \Arf[\rho + A] + n_2 \Arf[\rho]} \cZ_F[\Sigma, \rho, A]
\end{equation}
where $n_1, n_2$ are free parameters. This is because \eqref{eq:couplingseconddef} forces the freedom to be a sign, and compatibility with cutting and pasting requires these signs to form an SPT. Such SPTs are classified by ${(I_\bZ\Omega^\Spin})^3(B\bZ_2) = \bZ_2 \times \bZ_2$, with the two $\bZ_2$ factors corresponding to the choices $n_1$ and $n_2$ in \eqref{eq:freedom}.

Actually there is one further condition on $\cZ_F[\Sigma, \rho, A]$ we should impose. When the background $A$ is turned off, the coupled partition function should reduce to the original theory, allowing us to write the equation
\[
    \cZ_F[\Sigma, \rho, A = 0] = \cZ_F[\Sigma, \rho]
\]
Note that this time, both sides belong to the same Hilbert space $\cH_{\SO(n)_{-1}}[\Sigma, \rho]$, so this definition is unafflicted by the previous subtleties. This forces $n_1 = n_2$ in \eqref{eq:freedom}. Hence the final answer for the non-uniqueness is
\begin{equation}\label{eq:freedomreduced}
    \cZ_F[\Sigma, \rho, A] \rightarrow (-1)^{\Arf[\rho + A] + \Arf[\rho]} \cZ_F[\Sigma, \rho, A]
\end{equation}

We can summarise the above discussion as follows:

\begin{itemize}

\item When $n = 0$ mod 16, fermion parity is defined in the obvious fashion by \eqref{eq:couplingwrongdef},
\[
    \cZ_F[\Sigma, \rho, A] = \cZ_F[\Sigma, \rho + A]
\]

\item When $n = 8$ mod 16, fermion parity requires a more subtle definition given by the requirements
\begin{flalign*}
    &\cZ_F[\Sigma, \rho, A] \in \cH_{\SO(n)_{-1}}[\Sigma, \rho] \\
    &\cZ_F[\Sigma, \rho, A]^2 = \cZ_F[\Sigma, \rho + A]^2 \\
    &\cZ_F[\Sigma, \rho, A] \text{ obeys cutting and pasting} \\
    &\cZ_F[\Sigma, \rho, A = 0] = \cZ_F[\Sigma, \rho]
\end{flalign*}
There are two solutions related under \eqref{eq:freedomreduced}
\[
    \cZ_F[\Sigma, \rho, A] \rightarrow (-1)^{\Arf[\rho + A] + \Arf[\rho]} \cZ_F[\Sigma, \rho, A]
\]
neither of which can be canonically chosen over the other.

\end{itemize}

\subsection{Gauging Fermion Parity}\label{sec:gauging}

We would now like to briefly examine the implications for gauging. The definition of the gauged theory $T_{F'} = T_F / (-1)^F$ is
\begin{equation}\label{eq:gaugingdef2}
    \cZ_{F'}[\Sigma, \rho, A] = \frac{1}{\sqrt{|H^1(\Sigma; \bZ_2)|}} \sum_a (-1)^{\int_\Sigma a \smile A} \cZ_F[\Sigma, \rho, a]
\end{equation}
which is identical to \eqref{eq:gaugingdef} but for the replacement of $\cZ_F[\Sigma, \rho + a]$ with $\cZ_F[\Sigma, \rho, a]$. Note that all terms in the sum live in the same Hilbert space $\cH_{\SO(n)_{-1}}[\Sigma, \rho]$, so the sum makes sense.

When $n = 0$ mod 16, we can use the relation \eqref{eq:couplingwrongdef} to show that $T_{F'}$ is secretly a bosonic theory $T_B$ after supplementing with an appropriate counterterm, as in Section~\ref{sec:review}. But when $n = 8$ mod 16, no such relation holds, and the bosonic theory $T_B$ does not even exist. Instead, $T_{F'}$ is a genuinely fermionic theory, as evidenced by the fact its partition function
\[
    \cZ_{F'}[\Sigma, \rho, A] \in \cH_{\SO(n)_{-1}}[\Sigma, \rho]
\]
belongs to the Hilbert space of a spin-SPT. Another way to say this is that the gravitational anomaly of $T_{F'}$ is $n \CSgrav$, which is not the gravitational anomaly $16 n_B \CSgrav$ of any bosonic theory.

In summary, when $n = 8$ mod 16, we can define a ``refermionisation'' operation which takes an old fermionic theory $T_F$ to a new one $T_{F'}$ by the following recipe:
\begin{itemize}
\item Lift $\cZ_F[\Sigma, \rho]$ to $\cZ_F[\Sigma, \rho, A]$ in one of the two possible ways.
\item Gauge $A$ to get $\cZ_{F'}[\Sigma, \rho, A]$.
\item Set $A = 0$ again to get $\cZ_{F'}[\Sigma, \rho]$.
\end{itemize}
For all values of $n$ mod 16 other than 0 and 8, fermion parity is anomalous, and it does not make sense to gauge it.

\subsection*{Comments}

We close this section with some comments. First, due to the ambiguity in the lift, there are in fact two such refermionisation operations, neither of which is preferred over the other. If we denote them as $R^{(1)}$ and $R^{(2)}$, then by \eqref{eq:freedomreduced} they are related as
\begin{equation}\label{eq:Rrelation}
    R^{(1)} \Big[ \; T_F \; \Big] = R^{(2)} \Big[ \; T_F \times \Arf \; \Big] \times \Arf
\end{equation}
That is, they are conjugate under the operation of stacking with $\Arf$. Clearly this relation is completely symmetrical with respect to $R^{(1)}$ and $R^{(2)}$.

Second, the theory $T_{F'}$ also comes with a choice of $\bZ_2$ symmetry that we have chosen to ignore when we set $A = 0$. Unsurprisingly, it will turn out that this $\bZ_2$ is simply the $(-1)^F$ symmetry of $T_{F'}$, and so no information has been lost.

One can similarly discuss summing over spin structures and gauging fermion parity in 3d. Because the gravitational anomaly is always trivial $(I_\bZ \Omega^\Spin)^5(\pt) = 0$, the spin structure is anomaly-free. As a consequence, summing over spin structure and gauging fermion parity are equivalent (up to stacking a counterterm), analogous to the case $n = 0$ discussed at the beginning of Section \ref{sec:review}.

Finally, one can also define $\cZ_F[\Sigma, \rho, A]$ by introducing external data. One way is to define the partition function not only on the tuple $(\Sigma, \rho, A)$, but in the 2d-3d coupled system, with the metric, the spin structure and the $\bZ_2$ gauge field extended to the bulk. Another way is to pin the spin structure $\rho$ to be a \emph{reference spin structure} $\rho_0$, and define
\[
    \cZ_F[\Sigma, \rho_0, A] \coloneqq \cZ_F[\Sigma, \rho_0 + A]
\]
via a \emph{non-canonical} isomorphism. Here non-canonical means that after changing the reference $\rho_0 \to \rho_0'$, the partition function $\cZ_F[\Sigma, \rho_0', A]$ is no longer $\cZ_F[\Sigma, \rho_0' + A]$, but could inevitably acquire a sign. Such a scheme is tricky to extend to higher genus, but it can easily be carried out on the torus, and was the method adopted in \cite{8maj}. This also clarifies why summing over $A$ is allowed when $n = 8$ mod 16: the background is a fixed spin structure and unfixed $\bZ_2$ gauge field $A$, and the corresponding anomaly is simply $(\widetilde{I_\bZ\Omega^\Spin})^4(B\bZ_2) = \bZ_8$.

\section{Explicit CFT Calculations}\label{sec:cft}

Our goal in this section is to give a tour of the main points of Section~\ref{sec:anomalies} when $T_F$ is a conformal field theory on the torus $\Sigma = T^2$. In the place of abstract, formal arguments, we will give simple concrete calculations.

Recall that on the torus, there are four possible spin structures labelled by a choice of NS or R around the spatial and temporal circles. For each such spin structure $\rho$, one has a partition function $\cZ_F[\rho, \tau]$ where $\tau$ is the modular parameter of the torus, which plays the role of the metric.

The CFT regularisation makes a number of implicit choices for us that allow us to regard $\cZ_F[\rho, \tau]$ as a genuine complex number, with no mention of SPT Hilbert spaces in sight. But it comes at a cost: the $\cZ_F[\rho, \tau]$ transform projectively under modular transformations $\cS(\tau) = -1 / \tau$ and $\cT(\tau) = \tau + 1$. This is encoded by the diagram\footnote{A Majorana-Weyl fermion on the RR spin structure with a single fermion insertion has $\cZ = \eta$. This picks up $\sqrt{-i\tau}$ under $\cS$, with the $\sqrt{\tau}$ coming from the fermion, and the rest attributable to the partition function. This explains the phase $e^{-2 \pi i \frac{n}{8}}$.}
\begin{equation}\label{eq:anomalousphases}
    \hspace*{-3ex}
    \begin{tikzcd}
        \cZ_F\big[\tau,\spin{NS}{NS}\;\big] \arrow[blue, rightarrow, loop, in=160, out=200, distance=3em, "\cS"] \arrow[\greent, leftrightarrow, r, "\cT \textstyle = e^{-2 \pi i \frac{n}{48}}" {yshift=2ex}] &[+5ex]
        \cZ_F\big[\tau,\spin{R}{NS}\;\big] \arrow[blue, leftrightarrow, r, "\cS"] &
        \cZ_F\big[\tau,\spin{NS}{R}\;\big] \arrow[\greent, rightarrow, loop, distance=3em, in=340, out=20, "\cT \textstyle = e^{2 \pi i \frac{n}{24}}"] \\
        & & \cZ_F\big[\tau,\spin{R}{R}\;\big] \arrow[blue, rightarrow, loop, in=160, out=200, distance=3em, "\cS \textstyle = e^{-2 \pi i \frac{n}{8}}"] \arrow[\greent, rightarrow, loop, distance=3em, in=340, out=20, "\cT \textstyle = e^{2 \pi i \frac{n}{24}}"]
    \end{tikzcd}
\end{equation}
Here the notation $\cZ_F[\tau, \rho] \xrightarrow{\cT \textstyle = e^{i \theta}} \cZ_F[\tau, \rho']$, for example, means $\cZ_F[\tau + 1, \rho] = e^{i \theta} \cZ_F[\tau, \rho']$. These phases are the manifestation of the gravitational anomaly.

Since our interest lies in the case where we can gauge fermion parity, we will set $n = 8m$. Notice that this renders the anomalous phases of all $\cS$ transformations trivial. It will also be important for later that the remaining phases, belonging to the $\cT$ transformations, are related by a sign:
\begin{equation}\label{eq:signrel}
    e^{-2 \pi i \frac{m}{6}} = (-1)^m e^{2 \pi i \frac{m}{3}}
\end{equation}
The sign only manifests in the interesting case $n = 8$ mod 16, corresponding to $m$ odd.

\subsection{Coupling to Fermion Parity}\label{sec:cftcoupling}

We begin our tour with Section~\ref{sec:coupling}. As before, to warm up we quickly see why spin structure summation requires $n = 0$ mod 16. In order for the sum $\sum_\rho \cZ_F[\tau, \rho]$ to transform consistently, all $\cT$ transformations in \eqref{eq:anomalousphases} must have the same phase. This indeed requires $n = 0$ mod 16.

Next, we would like to couple $T_F$ to fermion parity. The obvious definition of the coupled partition function is $\cZ_F[\tau, \rho, A] \coloneqq \cZ_F[\tau, \rho + A]$. Let us denote the partition functions on the four possible spin structures by
\begin{alignat*}{2}
    \cZ_F\Big[\tau,\,\spin{NS}{NS}\;\Big] &= \secta(\tau) &\qquad \cZ_F\Big[\tau,\,\spin{R}{NS}\;\Big] &= \sectb(\tau) \\
    \cZ_F\Big[\tau,\,\spin{NS}{R}\;\Big] &= \sectc(\tau) &\qquad \cZ_F\Big[\tau,\,\spin{R}{R}\;\Big] &= \sectd(\tau)
\end{alignat*}
Then the coupled partition functions are also given by $\secta$, $\sectb$, $\sectc$ and $\sectd$. We have arranged these into the table shown in Figure~\ref{fig:bigtable1}, with rows labelled by $\rho$ and columns labelled by $A$. We have also annotated it with their modular transformation properties.

\begin{figure}
    \centering
    \begin{adjustbox}{scale=0.92,center}
        \begin{tikzcd}[sep=large,nodes in empty cells,execute at end picture={
            \draw
                ($(\tikzcdmatrixname-1-1.south)!0.5!(\tikzcdmatrixname-2-1.north)$) coordinate (lx)
                (\tikzcdmatrixname.west|-lx) -- (\tikzcdmatrixname.east|-lx);
            \draw
                ($(\tikzcdmatrixname-1-1.east)!0.5!(\tikzcdmatrixname-1-2.west)$) coordinate (ly)
                (ly|-\tikzcdmatrixname.north) -- (ly|-\tikzcdmatrixname.south);
            }]
            &
            \spin{}{} \arrow[\greent, dash, loop, in=170, out=190, distance=0.5em, "\cT"] \arrow[blue, dash, loop, in=-10, out=10, distance=0.5em, "\cS"] &
            \spinx{}{} \arrow[\greent, dash, loop, in=170, out=190, distance=0.5em, "\cT"] \arrow[blue, dash, r, "\cS"] &
            \spiny{}{} \arrow[\greent, dash, r, "\cT"] &
            \spinxy{}{} \arrow[blue, dash, loop, in=-10, out=10, distance=0.5em, "\cS"]
            \\
            \spin{NS}{NS}\hspace{1ex} \arrow[blue, dash, loop, in=80, out=100, distance=0.5em, "\cS"] \arrow[\greent, dash, d, "\cT"] &
            \secta \arrow[blue, dash, loop, in=80, out=100, distance=0.5em, "\cS"] \arrow[\greent, dash, d, "\cT \textstyle = e^{-2 \pi i \frac{m}{6}}"] &
            \sectb \arrow[blue, dash, r, "\cS"] \arrow[\greent, dash, d, "\cT \textstyle = e^{-2 \pi i \frac{m}{6}}"] &
            \sectc \arrow[\greent, dash, rd] &
            \sectd \arrow[blue, dash, loop, in=80, out=100, distance=0.5em, "\cS"] \arrow[\greent, phantom, d, "\scriptstyle \cT \textstyle = e^{2 \pi i \frac{m}{3}}"]
            \\
            \spin{R}{NS} \arrow[blue, dash, d, "\cS"] &
            \sectb \arrow[blue, dash, d, "\cS"] &
            \secta \arrow[blue, dash, rd] &
            \sectd \arrow[\greent, dash, ru] \arrow[blue, phantom, d, "\scriptstyle \cS" {xshift=-3.5ex}] &
            \sectc \arrow[blue, dash, d, "\cS"]
            \\
            \spin{NS}{R}\hspace{1ex} \arrow[\greent, dash, loop, in=260, out=280, distance=0.5em, "\cT"] &
            \sectc \arrow[\greent, dash, loop, in=260, out=280, distance=0.5em, "\cT \textstyle = e^{2 \pi i \frac{m}{3}}"] &
            \sectd \arrow[blue, dash, ru] \arrow[\greent, dash, loop, in=260, out=280, distance=0.5em, "\cT \textstyle = e^{2 \pi i \frac{m}{3}}"] &
            \secta \arrow[\greent, dash, r, swap, "\cT \textstyle = e^{-2 \pi i \frac{m}{6}}" {yshift=-2ex}] &
            \sectb
            \\
            \spin{R}{R} \arrow[blue, dash, loop, in=80, out=100, distance=0.5em, "\cS"] \arrow[\greent, dash, loop, in=260, out=280, distance=0.5em, "\cT"] &
            \sectd \arrow[blue, dash, loop, in=80, out=100, distance=0.5em, "\cS"] \arrow[\greent, dash, loop, in=260, out=280, distance=0.5em, "\cT \textstyle = e^{2 \pi i \frac{m}{3}}"] &
            \sectc \arrow[\greent, dash, loop, in=260, out=280, distance=0.5em, "\cT \textstyle = e^{2 \pi i \frac{m}{3}}"] \arrow[blue, dash, r, "\cS"] &
            \sectb \arrow[\greent, dash, r, swap, "\cT \textstyle = e^{-2 \pi i \frac{m}{6}}" {yshift=-2ex}] &
            \secta \arrow[blue, dash, loop, in=-10, out=10, distance=0.5em, "\cS"]
        \end{tikzcd}
    \end{adjustbox}
    \caption{The coupled partition function $\cZ_F[\tau, \rho, A]$ when $n = 0$ mod 16. \label{fig:bigtable1}}

    \vspace{3em}

    \begin{adjustbox}{scale=0.92,center}
        \begin{tikzcd}[sep=large,nodes in empty cells,execute at end picture={
            \draw
                ($(\tikzcdmatrixname-1-1.south)!0.5!(\tikzcdmatrixname-2-1.north)$) coordinate (lx)
                (\tikzcdmatrixname.west|-lx) -- (\tikzcdmatrixname.east|-lx);
            \draw
                ($(\tikzcdmatrixname-1-1.east)!0.5!(\tikzcdmatrixname-1-2.west)$) coordinate (ly)
                (ly|-\tikzcdmatrixname.north) -- (ly|-\tikzcdmatrixname.south);
            }]
            &
            \spin{}{} \arrow[\greent, dash, loop, in=170, out=190, distance=0.5em, "\cT"] \arrow[blue, dash, loop, in=-10, out=10, distance=0.5em, "\cS"] &
            \spinx{}{} \arrow[\greent, dash, loop, in=170, out=190, distance=0.5em, "\cT"] \arrow[blue, dash, r, "\cS"] &
            \spiny{}{} \arrow[\greent, dash, r, "\cT"] &
            \spinxy{}{} \arrow[blue, dash, loop, in=-10, out=10, distance=0.5em, "\cS"]
            \\
            \spin{NS}{NS}\hspace{1ex} \arrow[blue, dash, loop, in=80, out=100, distance=0.5em, "\cS"] \arrow[\greent, dash, d, "\cT"] &
            \secta \arrow[blue, dash, loop, in=80, out=100, distance=0.5em, "\cS"] \arrow[\greent, dash, d, "\cT \textstyle = e^{-2 \pi i \frac{m}{6}}"] &
            \sectb \arrow[blue, dash, r, "\cS"] \arrow[\greent, dash, d, "\cT \textstyle = e^{-2 \pi i \frac{m}{6}}"] &
            \sectc \arrow[\greent, dash, rd] &
            -\sectd \arrow[blue, dash, loop, in=80, out=100, distance=0.5em, "\cS"] \arrow[\greent, phantom, d, "\scriptstyle \cT \textstyle = e^{-2 \pi i \frac{m}{6}}" {xshift=0.5ex}]
            \\
            \spin{R}{NS} \arrow[blue, dash, d, "\cS"] &
            \sectb \arrow[blue, dash, d, "\cS"] &
            \secta \arrow[blue, dash, rd] &
            \sectd \arrow[\greent, dash, ru] \arrow[blue, phantom, d, "\scriptstyle \cS" {xshift=-3.5ex}] &
            -\sectc \arrow[blue, dash, d, "\cS"]
            \\
            \spin{NS}{R}\hspace{1ex} \arrow[\greent, dash, loop, in=260, out=280, distance=0.5em, "\cT"] &
            \sectc \arrow[\greent, dash, loop, in=260, out=280, distance=0.5em, "\cT \textstyle = e^{2 \pi i \frac{m}{3}}"] &
            \sectd \arrow[blue, dash, ru] \arrow[\greent, dash, loop, in=260, out=280, distance=0.5em, "\cT \textstyle = e^{2 \pi i \frac{m}{3}}"] &
            \secta \arrow[\greent, dash, r, swap, "\cT \textstyle = e^{2 \pi i \frac{m}{3}}" {yshift=-2ex}] &
            -\sectb
            \\
            \spin{R}{R} \arrow[blue, dash, loop, in=80, out=100, distance=0.5em, "\cS"] \arrow[\greent, dash, loop, in=260, out=280, distance=0.5em, "\cT"] &
            \sectd \arrow[blue, dash, loop, in=80, out=100, distance=0.5em, "\cS"] \arrow[\greent, dash, loop, in=260, out=280, distance=0.5em, "\cT \textstyle = e^{2 \pi i \frac{m}{3}}"] &
            \sectc \arrow[\greent, dash, loop, in=260, out=280, distance=0.5em, "\cT \textstyle = e^{2 \pi i \frac{m}{3}}"] \arrow[blue, dash, r, "\cS"] &
            \sectb \arrow[\greent, dash, r, swap, "\cT \textstyle = e^{2 \pi i \frac{m}{3}}" {yshift=-2ex}] &
            -\secta \arrow[blue, dash, loop, in=-10, out=10, distance=0.5em, "\cS"]
        \end{tikzcd}
    \end{adjustbox}
    \caption{One possible coupled partition function $\cZ_F[\tau, \rho, A]$ when $n = 8$ mod 16. \label{fig:bigtable2}}
\end{figure}

Recall that our goal is to have an \emph{anomaly-free} regularisation of $(-1)^F$. This means that the anomalous phases in Figure~\ref{fig:bigtable1} must be \emph{independent of $A$}, that is, constant across each row. We see that by equation~\eqref{eq:signrel}, this is only true for $m$ even. Thus the obvious definition of the coupled partition function is only valid for $n = 0$ mod 16, which parallels our discussion from earlier.

However, there should also exist an anomaly-free regularisation of $(-1)^F$ when $n = 8$ mod 16. This can be achieved by flipping signs in Figure~\ref{fig:bigtable1} to ensure that the anomalous phases are constant across each row. In Figure~\ref{fig:bigtable2}, we have made one such choice. This is therefore the definition of $\cZ_F[\tau, \rho, A]$ we must use for $n = 8$ mod 16.

The solution in Figure~\ref{fig:bigtable2} is by no means the only one. There are five disconnected orbits under modular transformations, and each can seemingly be flipped independently, giving $2^5$ solutions. But between the fact the $A = 0$ column must stay fixed, the $\spiny{NS}{NS}$ entry must stay positive, and the fact that adding a $\spinx{}{}$ line must commute with flipping the spin structure around the time cycle, the only ambiguity in the solution is
\begin{equation}\label{eq:sptpattern}
    \def\arraystretch{1.4}
    \arraycolsep=4pt
    \begin{array}{r|cccc}
        & \spin{}{} & \spinx{}{} & \spiny{}{} & \spinxy{}{} \\
        \hline
        \spin{NS}{NS} & + & + & + & - \\
        \spin{R}{NS} & + & + & - & + \\
        \spin{NS}{R} & + & - & + & + \\
        \spin{R}{R} & + & - & - & -
    \end{array}
\end{equation}
This pattern of signs is precisely the SPT phase $(-1)^{\Arf[\rho + A] + \Arf[\rho]}$. Thus $\cZ_F[\tau, \rho, A]$ is unique up to stacking with this SPT, as claimed in \eqref{eq:freedomreduced}.

\subsection{Gauging Fermion Parity}\label{sec:cftgauging}

Now let's see what happens when we gauge $(-1)^F$ in Figures~\ref{fig:bigtable1} and \ref{fig:bigtable2}. This will produce a new theory $T_{F'} = T_F / (-1)^F$, whose partition function $\cZ_{F'}[\tau, \rho, A]$ we will calculate.

We begin with the simpler case $n = 0$ mod 16. Using \eqref{eq:gaugingdef2}, gauging $(-1)^F$ acts on Figure~\ref{fig:bigtable1} as follows:
\[
    \def\arraystretch{1.4}
    \arraycolsep=4pt
    \begin{array}{r|cccc}
        & \spin{}{} & \spinx{}{} & \spiny{}{} & \spinxy{}{} \\
        \hline
        \spin{NS}{NS} & \secta & \sectb & \sectc & \sectd \\
        \spin{R}{NS} & \sectb & \secta & \sectd & \sectc \\
        \spin{NS}{R} & \sectc & \sectd & \secta & \sectb \\
        \spin{R}{R} & \sectd & \sectc & \sectb & \secta
    \end{array}
    \qquad \rightarrow \qquad
    \begin{array}{r|rrrr}
        & \spin{}{} & \spinx{}{} & \spiny{}{} & \spinxy{}{} \\
        \hline
        \spin{NS}{NS} & \secta' & \sectb' & \sectc' & -\sectd' \\
        \spin{R}{NS} & \secta' & \sectb' & -\sectc' & \sectd' \\
        \spin{NS}{R} & \secta' & -\sectb' & \sectc' & \sectd' \\
        \spin{R}{R} & \secta' & -\sectb' & -\sectc' & -\sectd'
    \end{array}
\]
The right hand side is the partition function $\cZ_{F'}[\tau, \rho, A]$, with entries given by
\begin{align*}
    \secta' &= \frac{\secta + \sectb + \sectc + \sectd}{2} & \sectb' &= \frac{\secta + \sectb - \sectc - \sectd}{2} \\
    \sectc' &= \frac{\secta - \sectb + \sectc - \sectd}{2} & \sectd' &= \frac{-\secta + \sectb + \sectc - \sectd}{2}
\end{align*}
Entirely as expected, the resulting partition function is independent of spin structure up to an overall topological counterterm $(-1)^{\Arf[\rho + A] + \Arf[\rho]}$ that we recognise from \eqref{eq:sptpattern}. Stripping it off, we find a bosonic theory $T_B$ whose partition functions $\cZ_B[\tau, A]$ are given by $\secta'$, $\sectb'$, $\sectc'$ and $\sectd'$. This is the standard bosonisation story we reviewed in Section~\ref{sec:review}.

How about the more interesting case $n = 8$ mod 16? This time gauging $(-1)^F$ acts on Figure~\ref{fig:bigtable2} as
\[
    \def\arraystretch{1.4}
    \arraycolsep=4pt
    \begin{array}{r|rrrr}
        & \spin{}{} & \spinx{}{} & \spiny{}{} & \spinxy{}{} \\
        \hline
        \spin{NS}{NS} & \secta & \sectb & \sectc & -\sectd \\
        \spin{R}{NS} & \sectb & \secta & \sectd & -\sectc \\
        \spin{NS}{R} & \sectc & \sectd & \secta & -\sectb \\
        \spin{R}{R} & \sectd & \sectc & \sectb & -\secta
    \end{array}
    \qquad \rightarrow \qquad
    \begin{array}{r|cccc}
        & \spin{}{} & \spinx{}{} & \spiny{}{} & \spinxy{}{} \\
        \hline
        \spin{NS}{NS} & \secta' & \sectb' & \sectc' & -\sectd' \\
        \spin{R}{NS} & \sectb' & \secta' & \sectd' & -\sectc' \\
        \spin{NS}{R} & \sectc' & \sectd' & \secta' & -\sectb' \\
        \spin{R}{R} & \sectd' & \sectc' & \sectb' & -\secta'
    \end{array}
\]
with entries given by
\begin{equation}\label{eq:elusive_matrix_elements}
    \begin{aligned}
        \secta' &= \frac{\secta + \sectb + \sectc - \sectd}{2} &\qquad\qquad \sectb' &= \frac{\secta + \sectb - \sectc + \sectd}{2} \\
        \sectc' &= \frac{\secta - \sectb + \sectc + \sectd}{2} &\qquad\qquad \sectd' &= \frac{-\secta + \sectb + \sectc + \sectd}{2}
    \end{aligned}
\end{equation}
But this is another copy of the same table. It is therefore to be interpreted as a new genuinely fermionic theory $\cZ_{F'}[\rho]$. This explicitly illustrates the mechanism by which gauging $(-1)^F$ does not eliminate the spin structure dependence.

Note that in Section~\ref{sec:anomalies} we asserted that when $n = 8$ mod 16, the extra $\bZ_2$ symmetry of $\cZ_{F'}[\tau, \rho, A]$ was simply the fermion parity symmetry of $T_{F'}$. Here we see this claim borne out, by the fact the second table is the same as the first table, but with dashes.

It's also easy to demonstrate the relation \eqref{eq:Rrelation} between the two refermionisation operations $R^{(1)}$ and $R^{(2)}$; all we have to do is stack Figure~\ref{fig:bigtable2} with the pattern of phases \eqref{eq:sptpattern} and redo the gauging calculation that lead to \eqref{eq:elusive_matrix_elements}.

\subsection{Duality Webs}

As commented in Section \ref{sec:review}, in addition to bosonisation (for $n = 0$ mod 16) and refermionisation (for $n = 8$ mod 16), for any $n$, one can also stack an Arf invariant to map one theory to another theory. Combining them together, the various theories form a duality web, which we analyse below.

When $n = 0$ mod 16, the bosonisation (Bos) map, the fermionisation (Fer) map, and stacking an Arf invariant relate four theories,
\begin{equation}
    \begin{tikzpicture}
        \node[] at (0,0) {$T_F$};
        \node[] at (4,0) {$T_F \times \Arf$};
        \node[] at (0,-3) {$T_B$};
        \node[] at (4,-3) {$T_B / \bZ_2$};
        \draw[red, <->] (0.5,0) -- (3,0);
        \node[red, above] at (1.75,0) {\footnotesize{$\times \Arf$}};
        \draw[orange, <->] (0.5,-3) -- (3,-3);
        \node[orange, above] at (1.75,-3) {\footnotesize{$/\bZ_2$}};
        \draw[blue, ->] (-0.1,-0.5) -- (-0.1,-2.5);
        \node[blue, left] at (-0.1,-1.5) {\footnotesize{Bos}};
        \draw[blue, <-] (0.1,-0.5) -- (0.1,-2.5);
        \node[blue, right] at (0.1,-1.5) {\footnotesize{Fer}};
        \draw[blue, ->] (3.9,-0.5) -- (3.9,-2.5);
        \node[blue, left] at (3.9,-1.5) {\footnotesize{Bos}};
        \draw[blue, <-] (4.1,-0.5) -- (4.1,-2.5);
        \node[blue, right] at (4.1,-1.5) {\footnotesize{Fer}};
    \end{tikzpicture}
\end{equation}
The bosonisation map is defined in \eqref{eq:spinsumdef}, and the fermionisation map is its inverse. The $/\bZ_2$ is gauging the $\bZ_2$ symmetry of the bosonic theory. The $\times \Arf$ maps $\cZ_F[\Sigma, \rho] \rightarrow \cZ_F[\Sigma, \rho] (-1)^{\Arf[\rho]}$; consequently $\cZ_F[\Sigma, \rho, A] \rightarrow \cZ_F[\Sigma, \rho, A] (-1)^{\Arf[\rho + A]}$. This duality web has been extensively studied in recent years \cite{Ji:2019ugf, Karch:2019lnn, Hsieh:2020uwb}.

When $n=8$ mod 16, there are two refermionisation operations $R^{(1)}$ and $R^{(2)}$. The CFT regularisation allows us to privilege one over the other; for example, we can identify $R^{(1)}$ with the operation $(\secta, \sectb, \sectc, \sectd) \rightarrow (\secta', \sectb', \sectc', \sectd')$ defined by \eqref{eq:elusive_matrix_elements}, and $R^{(2)}$ with the other operation, related by \eqref{eq:Rrelation}. Then depending on whether we choose to define the gauged theory $T_{F'}$ using $R^{(1)}$ or $R^{(2)}$, the effect of gauging $(-1)^F$ is to permute the twisted sectors in the following way:
\[
    \begin{tikzcd}[ampersand replacement=\&, column sep=huge, row sep=0]
        \&
        T_{F'} \, : \quad
        \begin{array}{c|cc}
            & + & - \\
            \hline
            \NS & \cellcolor[HTML]{\cola} \frac{\secta+\sectb}{2} & \cellcolor[HTML]{\cold} \frac{\sectc-\sectd}{2} \\
            \R  & \cellcolor[HTML]{\colc} \frac{\sectc+\sectd}{2} & \cellcolor[HTML]{\colb} \frac{\secta-\sectb}{2}
        \end{array}
        \\
        T_F \, : \quad
        \begin{array}{c|cc}
            & + & - \\
            \hline
            \NS & \cellcolor[HTML]{\cola} \frac{\secta+\sectb}{2} & \cellcolor[HTML]{\colb} \frac{\secta-\sectb}{2} \\
            \R  & \cellcolor[HTML]{\colc} \frac{\sectc+\sectd}{2} & \cellcolor[HTML]{\cold} \frac{\sectc-\sectd}{2}
        \end{array}
        \arrow[rightarrow, ru, "R^{(1)}"]
        \arrow[rightarrow, rd, "R^{(2)}"]
        \& \\
        \&
        T_{F'} \, : \quad
        \begin{array}{c|cc}
            & + & - \\
            \hline
            \NS & \cellcolor[HTML]{\cola} \frac{\secta+\sectb}{2} & \cellcolor[HTML]{\colc} \frac{\sectc+\sectd}{2} \\
            \R  & \cellcolor[HTML]{\colb} \frac{\secta-\sectb}{2} & \cellcolor[HTML]{\cold} \frac{\sectc-\sectd}{2}
        \end{array}
    \end{tikzcd}
\]
Both of these are evidently order-2 operations on the space of fermionic quantum field theories, as would be expected from gauging a $\bZ_2$ symmetry. The operation of stacking with $\Arf$ is also of order 2, and exchanges the two entries in the bottom row. Together these operations form the group $\bS_3$, and fit together into a duality web that looks like the Cayley graph for $\bS_3$,
\begin{equation}\label{eq:dualityweb}
    \begin{tikzcd}[column sep=large, row sep=large]
        & T_{F_1} \times \Arf \arrow[r, blue, leftrightarrow, "/(-1)^F"] & T_{F_2} \arrow[dr, red, leftrightarrow, "\times \Arf"] & \\
        T_{F_1} \arrow[ur, red, leftrightarrow, "\times \Arf"] & & & T_{F_2} \times \Arf \arrow[dl, blue, leftrightarrow, "/(-1)^F"] \\
        & T_{F_3} \times \Arf \arrow[ul, blue, leftrightarrow, "/(-1)^F"] & T_{F_3} \arrow[l, red, leftrightarrow, "\times \Arf"] &
    \end{tikzcd}
\end{equation}
where $/(-1)^F$ denotes a choice of either $R^{(1)}$ or $R^{(2)}$, the same for all edges. This diagram would have been slightly tricky to derive using the formalism of Section~\ref{sec:anomalies}. However our current derivation is limited to CFTs on the torus. In the next section we will see that it continues to hold in the general case too, using the formalism of Symmetry TFT.

Finally, note that a very similar structure to \eqref{eq:dualityweb} also exists for fermionic theories with a specified choice of $\bZ_2$ symmetry and any gravitational anomaly \cite{bhardwaj2020}. We comment more on their precise relation in Appendix~\ref{app:web}.

\section{Symmetry TFT}\label{sec:symtft}

As discussed in Section~\ref{sec:grav}, the partition function of any 2d fermionic theory $T_F$ with gravitational anomaly $n$ can be viewed as a state in the Hilbert space of $\SO(n)_{-1}$. We can denote this bulk-boundary relation diagrammatically as
\[
    \begin{tikzpicture}[baseline={([yshift=-.5ex](0,1.5))}]
        \shade[top color=dgreen!30, bottom color=dgreen!5] (0,0) rectangle (4,3);
        \draw[line width=1pt] (4,0) -- (4,3);
        \node at (2,1.5) {$\SO(n)_{-1}$};
        \node[below] at (4,0) {$\cZ_F[\rho]$};
    \end{tikzpicture}
    \qquad \iff \qquad
    \cZ_F[\rho] \in \cH_{\SO(n)_{-1}}[\rho]
\]
where we suppress the metric $g$ and manifold $\Sigma$ dependence for simplicity.

It is useful to sum the bulk over spin structures. Unlike in 2d, this is always possible since there are no gravitational anomalies in 3d, and the result is $\Spin(n)_{-1}$. This is a bosonic TFT. It governs the topological manipulations \cite{orbifoldgroupoids} one can perform involving the spin structure, and is known as the \emph{Symmetry TFT} (SymTFT) for the spin structure dependence of $T_F$. The Symmetry TFT should know everything about summing over spin structures and gauging $(-1)^F$ that we have discussed in the previous sections. Our goal in this section is to show how.

\subsection{Properties of the Symmetry TFT}\label{sec:propSymTFT}

We begin by elucidating the relation between the SymTFT and the theory $T_F$ in a little more detail. By construction, we have the relation of partition functions
\begin{equation}\label{eq:sumSPT}
    \cZ_{\Spin(n)_{-1}} = \sum_{\rho_3} \cZ_{\SO(n)_{-1}}[\rho_3]
\end{equation}
on a fixed closed 3-manifold. This implies the relation among Hilbert spaces
\begin{equation}\label{eq:sumH}
    \cH_{\Spin(n)_{-1}} = \bigoplus_\rho \cH_{\SO(n)_{-1}}[\rho]
\end{equation}
on any 2d surface. This is interesting. It says that although we cannot sum the states $\cZ_F[\rho]$, a point we belaboured in Section~\ref{sec:anomalies}, we can \emph{formally} sum them, and the result will be a state in the Hilbert space of $\Spin(n)_{-1}$. We denote this state by
\[
    \ket{\cZ_F} \coloneqq \sum_\rho \cZ_F[\rho] \in \cH_{\Spin(n)_{-1}}
\]
The state $\ket{\cZ_F}$ encodes the response of $T_F$ to all spin structures at the same time. It also defines a boundary condition for $\Spin(n)_{-1}$. Its relation to the original partition function is encoded by the diagrammatic equation
\[
    \begin{tikzpicture}
        \shade[top color=dgreen!30, bottom color=dgreen!5] (0,0) rectangle (3,3);
        \shade[top color=blue!30, bottom color=blue!5] (3,0) rectangle (6,3);
        \shade[top color=dgreen!30, bottom color=dgreen!5] (9,0) rectangle (13,3);
        \draw[line width=1pt] (3,0) -- (3,3);
        \draw[line width=1pt] (6,0) -- (6,3);
        \draw[line width=1pt] (13,0) -- (13,3);
        \node at (1.5,1.5) {$\SO(n)_{-1}$};
        \node at (4.5,1.5) {$\Spin(n)_{-1}$};
        \node at (7.5,1.5) {$=$};
        \node at (11,1.5) {$\SO(n)_{-1}$};
        \node[below] at (3,0) {$\cI_{V}[\rho]$};
        \node[below] at (6,0) {$\ket{\cZ_F}$};
        \node[below] at (13,0) {$\cZ_F[\rho]$};
    \end{tikzpicture}
\]
Here we have introduced the interface $\cI_{V}[\rho]$, defined as the projector onto the $\rho$th factor of \eqref{eq:sumH}. The meaning of the subscript $V$ will become clear below. $\cI_V[\rho]$ can alternatively be written as $\ket{\rho} \bra{\rho}$, where $\ket{\rho}$ is a choice of basis state for $\cH_{\SO(n)_{-1}}[\rho]$, and $\bra{\rho}$ is the same state regarded as an element of $\cH_{\Spin(n)_{-1}}^*$ via \eqref{eq:sumH}. A crucial point is that due to the anomaly of the spin structure, both $\ket{\rho}$ and $\bra{\rho}$ are ambiguous up to a phase. However, the ambiguities cancel against each other in $\ket{\rho} \bra{\rho}$, hence $\cI_V[\rho]$ defines an unambiguous topological interface. For this reason, we use $\cI_V[\rho]$ throughout this section. Physically, we can think of $\Spin(n)_{-1}$ as a theory with a dynamical spin structure $\rho_3$, and $\cI_V[\rho]$ as a Dirichlet boundary condition for $\rho_3$ on the interface. As depicted above, fusing this interface with $\ket{\cZ_F}$ yields the original partition function $\cZ_F[\rho]$, replete with attached SPT.

\subsubsection*{Line Operators}

We'll need some basic properties of the line operators of $\Spin(n)_{-1}$ \cite{Seiberg:2016rsg}. These are classified by the representations of $\Spin(n)$ of level $\leq 1$, namely the trivial (1), vector ($V$), spinor ($S$) and for $n$ even the conjugate spinor ($C$) representations. Their topological spins are
\[
    \arraycolsep=5pt
    \def\arraystretch{2}
    \begin{array}{c|cccc}
    \text{line} & 1 & V & S & C \\
    \hline
    \text{spin} & 0 & \dfrac{1}{2} & \dfrac{n}{16} & \dfrac{n}{16}
    \end{array}
\]
Unless $n = 0$ mod 16, none of them can be condensed to produce a bosonic invertible theory. However, one can always condense $V$ to produce a fermionic invertible theory, dubbed fermionic anyon condensation \cite{Gaiotto:2015zta, Aasen:2017ubm}. Condensing $V$ on a half-space produces the interface $\cI_V[\rho]$, hence justifying the subscript of $\cI_V$.

\subsubsection*{The Arf Interface}

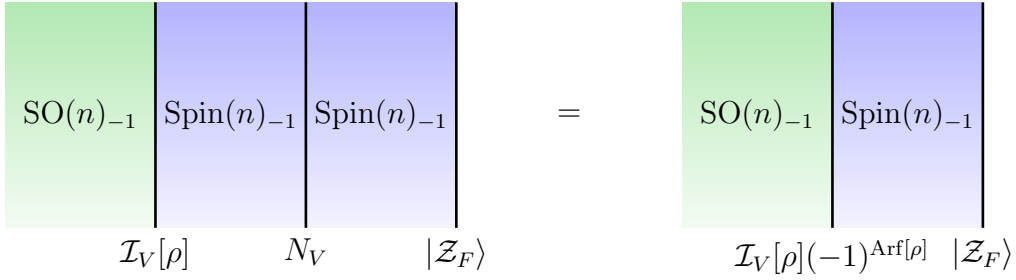
\begin{figure}
    \centering
    \begin{tikzpicture}
        \shade[top color=dgreen!30, bottom color=dgreen!5] (0,0) rectangle (2,3);
        \shade[top color=blue!30, bottom color=blue!5] (2,0) rectangle (6,3);
        \shade[top color=dgreen!30, bottom color=dgreen!5] (9,0) rectangle (11,3);
        \shade[top color=blue!30, bottom color=blue!5] (11,0) rectangle (13,3);
        \draw[line width=1pt] (2,0) -- (2,3);
        \draw[line width=1pt] (4,0) -- (4,3);
        \draw[line width=1pt] (6,0) -- (6,3);
        \draw[line width=1pt] (11,0) -- (11,3);
        \draw[line width=1pt] (13,0) -- (13,3);
        \node at (1,1.5) {$\SO(n)_{-1}$};
        \node at (3,1.5) {$\Spin(n)_{-1}$};
        \node at (5,1.5) {$\Spin(n)_{-1}$};
        \node at (7.5,1.5) {$=$};
        \node at (10,1.5) {$\SO(n)_{-1}$};
        \node at (12,1.5) {$\Spin(n)_{-1}$};
        \node[below] at (2,0) {$\cI_V[\rho]$};
        \node[below] at (4,0) {$N_V$};
        \node[below] at (6,0) {$\ket{\cZ_F}$};
        \node[below] at (11,0) {$\cI_V[\rho](-1)^{\Arf[\rho]}$};
        \node[below] at (13,0) {$\ket{\cZ_F}$};
    \end{tikzpicture}
    \caption{Insertion of $N_V$ implements stacking with $\Arf$.}
    \label{fig:stacking}
\end{figure}

We can already see some payoff from our setup by considering the topological operation $T_F \rightarrow T_F \times \Arf$. This can be performed for any value of the gravitational anomaly, and this fact should be reflected in some feature of $\Spin(n)_{-1}$ present for all $n$. To see how, we define a surface operator $N_V(\Sigma)$ in $\Spin(n)_{-1}$ by
\begin{equation}\label{eq:Nv}
    N_V(\Sigma) = (-1)^{\Arf[\rho_3 |_\Sigma]}
\end{equation}
where $\rho_3$ is the dynamical spin structure in the bulk, and $\rho_3 |_\Sigma$ is the induced spin structure on $\Sigma$. This is an invertible, order-2 defect. Its action on Hilbert space is $N_V \ket{\rho} = (-1)^{\Arf[\rho]} \ket{\rho}$. It follows that inserting $N_V$ into the bulk of the SymTFT implements the operation of stacking with Arf, as shown in Figure~\ref{fig:stacking}. From now on we will suppress the manifold $\Sigma$ dependence for the surface operator, unless necessary.

The surface $N_V$ can be constructed as the condensation defect for $V$,
\begin{equation}\label{eq:NV}
    N_V = \frac{1}{\sqrt{|H^1(\Sigma; \bZ_2)|}} \sum_{\gamma \in H_1(\Sigma; \bZ_2)} L_V(\gamma)
\end{equation}
Since $V$ is always fermionic for any value of $n$, condensing it  results in an invertible defect implementing a $\bZ_2$ symmetry \cite{Roumpedakis:2022aik, Buican:2023bzl}. To see how this defect acts on the interface $\cI_V[\rho]$, we first note that the fermionic line $L_V(\gamma)$ acts on the interface from the right by
\begin{equation}\label{eq:IVLV}
    \cI_V[\rho] L_V(\gamma) = \cI_V[\rho] (-1)^{\Arf[\rho + \PD(\gamma)] + \Arf[\rho]}
\end{equation}
There are two Arf factors on the right hand side because we require that the line operator acts trivially when $\gamma$ vanishes. Using the identities of the Arf invariant (see Appendix \ref{app:arf}), we find that $N_V$ acts on $\cI_V[\rho]$ as multiplication by a phase $(-1)^{\Arf[\rho]}$, as claimed in \eqref{eq:Nv}.

In what follows, we will analyse spin structure summation and gauging $(-1)^F$ in terms of the other condensation defects.

\subsection{The Case $n = 0$ mod 16}\label{sec:bos_tft}

The Symmetry TFT perspective on bosonisation was already discussed for $n = 0$ in \cite{ytlecture, orbifoldgroupoids}. Little changes when we generalise to $n = 0$ mod 16. However, we will still give a brief review to warm up, and as a foil to the discussion of the $n = 8$ mod 16 case. Throughout this section, we take $n = 16n_B$.

\subsubsection*{Interfaces between SymTFT and Invertible Theories}

When $n = 16n_B$, both the $S$ and $C$ lines are bosonic. Both are therefore condensable; condensing them produces the bosonic SPT $\Spin(16n_B)_{-1} / \bZ_2^S$.\footnote{The 3d theories $\Spin(16n_B)_{-1} / \bZ_2^S$ and $\Spin(16n_B)_{-1} / \bZ_2^C$ are the same bosonic invertible theory, and we use the former notation to represent both.} We will denote the interfaces associated with condensing $L_S$ and $L_C$ as $\cI_S$ and $\cI_C$ respectively. For $\cI_S$, $L_C$ is unscreened, and activates a $\bZ_2$ symmetry background field $A$. Similarly for $\cI_C$. Altogether, there are three interfaces we are interested in: $\cI_S[A]$, $\cI_C[A]$ and $\cI_V[\rho]$. The first two are genuine topological interfaces, while the third one is an almost topological interface that depends on the spin structure.

How do the lines $L_S, L_C, L_V$ act on the interfaces from the right? To determine this, we note that $L_S$ condenses on the $\cI_S$ boundary, or more precisely, $\cI_S[0] L_S(\gamma) = \cI_S[0]$. Turning on the background gauge field on the interface $\cI_S[A]$, it is natural to require
\begin{equation}
    \cI_S[A] L_S(\gamma) = \cI_S[A] (-1)^{\int_\gamma A}
\end{equation}
Similarly, the $L_C$ line acts on the $\cI_C[A]$ interface as
\begin{equation}
    \cI_C[A] L_C(\gamma) = \cI_C[A] (-1)^{\int_\gamma A}
\end{equation}
Together with \eqref{eq:IVLV}, and the commutation relation\footnote{See for instance \cite{Kaidi:2022cpf, Roumpedakis:2022aik} for recent discussions.}
\begin{equation}\label{eq:comm}
    L_S(\gamma) L_C(\gamma') = (-1)^{\braket{\gamma, \gamma'}} L_C(\gamma') L_S(\gamma)
\end{equation}
we are able to determine how an arbitrary line $L_{S,C,V}$ acts on an arbitrary interface $\cI_{S,C,V}$ from the right. We enumerate the results below:
\begin{equation}\label{eq:IVLS}
    \begin{split}
        \cI_S[A] L_S(\gamma) &= \cI_S[A] (-1)^{\int_\gamma A} \\
        \cI_S[A] L_C(\gamma) &= \cI_S[A + \PD(\gamma)] \\
        \cI_S[A] L_V(\gamma) &= \cI_S[A + \PD(\gamma)] (-1)^{\int_\gamma A} \\
        \cI_C[A] L_S(\gamma) &= \cI_C[A + \PD(\gamma)] \\
        \cI_C[A] L_C(\gamma) &= \cI_C[A] (-1)^{\int_\gamma A} \\
        \cI_C[A] L_V(\gamma) &= \cI_C[A + \PD(\gamma)] (-1)^{\int_\gamma A} \\
        \cI_V[\rho] L_S(\gamma) &= \cI_V[\rho + \PD(\gamma)] \\
        \cI_V[\rho] L_C(\gamma) &= \cI_V[\rho + \PD(\gamma)] (-1)^{\Arf[\rho + \PD(\gamma)] + \Arf[\rho]} \\
        \cI_V[\rho] L_V(\gamma) &= \cI_V[\rho] (-1)^{\Arf[\rho + \PD(\gamma)] + \Arf[\rho]}
    \end{split}
\end{equation}
The 7th and 8th equalities in \eqref{eq:IVLS} deserve additional comment. Consider the 7th equality in \eqref{eq:IVLS}: the left hand side is in the Hilbert space $\cH_{\SO(16n_B)_{-1}}[\rho] \otimes \cH_{\Spin(16n_B)_{-1}}^*$, while the right hand side is in the Hilbert space $\cH_{\SO(16n_B)_{-1}}[\rho + \PD(\gamma)] \otimes \cH_{\Spin(16n_B)_{-1}}^*$. Since the anomaly of the spin structure vanishes, $\cH_{\SO(n)_{-1}}[\rho]$ is canonically isomorphic to $\cH_{\SO(n)_{-1}}[\rho + \PD(\gamma)]$, hence the Hilbert spaces on both sides are canonically isomorphic. The above conditions are consistent with the relations between the interfaces
\begin{equation}\label{eq:interfacerelations}
    \begin{split}
        \cI_C[A] &= \frac{1}{\sqrt{|H^1(\Sigma; \bZ_2)|}} \sum_{a \in H^1(\Sigma; \bZ_2)} (-1)^{\int_\Sigma A \smile a} \cI_S[a] \\
        \cI_V[\rho] &= \frac{1}{\sqrt{|H^1(\Sigma; \bZ_2)|}} \sum_{a \in H^1(\Sigma; \bZ_2)} (-1)^{\Arf[\rho + a] + \Arf[\rho]} \cI_S[a]
    \end{split}
\end{equation}
The second identity makes sense because there is a canonical isomorphism between $\cH_{\SO(n)_{-1}}[\rho]$ and $\cH_{\Spin(n)_{-1}/\bZ_2^S}$  for $16|n$.

\subsubsection*{Invertible Defects in SymTFT}

To see how the above interfaces are related to each other, we enumerate the invertible topological defects. In Section \ref{sec:propSymTFT}, we already discussed one, $N_V$, obtained by condensing $L_V$ on $\Sigma$ \cite{Roumpedakis:2022aik}. There are two more, obtained by condensing the bosonic lines $L_S$ and $L_C$ respectively, weighted by an Arf dependent phase,
\begin{equation}\label{eq:NSNC}
    \begin{split}
        N_S[\rho] &= \frac{1}{\sqrt{|H^1(\Sigma; \bZ_2)|}} \sum_{\gamma \in H_1(\Sigma; \bZ_2)} (-1)^{\Arf[\rho + \PD(\gamma)]} L_S(\gamma) \\
        N_C[\rho] &= \frac{1}{\sqrt{|H^1(\Sigma; \bZ_2)|}} \sum_{\gamma \in H_1(\Sigma; \bZ_2)} (-1)^{\Arf[\rho + \PD(\gamma)]} L_C(\gamma)
    \end{split}
\end{equation}
Using \eqref{eq:comm} and the identities of Arf invariants, it is straightforward to check all three operators $N_S[\rho], N_C[\rho], N_V$ obey the $\bZ_2$ fusion rule (hence are invertible)
\begin{equation}
    N_S[\rho]^2 = N_C[\rho]^2 = N_V^2 = 1
\end{equation}
and $N_S[\rho]$ and $N_C[\rho]$ are mapped to each other under conjugation by $N_V$,
\begin{equation}
    N_C[\rho] = N_V N_S[\rho] N_V, \qquad N_S[\rho] = N_V N_C[\rho] N_V
\end{equation}

The invertible defects $N_{S,C}[\rho]$ depend upon a choice of spin structure $\rho$. By most reasonable definitions, $N_{S,C}[\rho]$ is a topological operator, since it is invariant under continuous deformations. But due to the choice of spin structure, $N_{S,C}[\rho]$ fails to be invariant under large deformations that return $\Sigma$ to itself up to a large diffeomorphism that shifts the spin structure. It is therefore a weaker notion of topological operator than the usual one (which only requires an orientation).  We will not view such an operator as strictly topological, but \emph{almost} topological.

These invertible surface operators obey the following commutation relations with the line operators:
\begin{equation}
    \begin{split}
        N_V L_S(\gamma) &= L_C(\gamma) N_V \\
        N_V L_C(\gamma) &= L_S(\gamma) N_V \\
        N_V L_V(\gamma) &= L_V(\gamma) N_V \\
        N_S[\rho] L_S(\gamma) &= L_S(\gamma) N_S[\rho] \\
        N_S[\rho] L_C(\gamma) &= L_V(\gamma) N_S[\rho] (-1)^{\Arf[\rho + \PD(\gamma)] + \Arf[\rho]} \\
        N_S[\rho] L_V(\gamma) &= L_C(\gamma) N_S[\rho] (-1)^{\Arf[\rho + \PD(\gamma)] + \Arf[\rho]} \\
        N_C[\rho] L_S(\gamma) &= L_V(\gamma) N_C[\rho] (-1)^{\Arf[\rho + \PD(\gamma)] + \Arf[\rho]} \\
        N_C[\rho] L_C(\gamma) &= L_C(\gamma) N_C[\rho] \\
        N_C[\rho] L_V(\gamma) &= L_S(\gamma) N_C[\rho] (-1)^{\Arf[\rho + \PD(\gamma)] + \Arf[\rho]}
    \end{split}
\end{equation}
Combining the above with the definition of surface operators, we finally arrive at the action of surface operators on the interfaces,
\begin{equation}\label{eq:operinterface1}
    \begin{split}
        \cI_S[A] N_V &= \cI_C[A] \\
        \cI_C[A] N_V &= \cI_S[A] \\
        \cI_V[\rho] N_V &= \cI_V[\rho] (-1)^{\Arf[\rho]} \\
        \cI_S[A] N_S[\rho] &= \cI_S[A] (-1)^{\Arf[\rho + A] + \Arf[\rho]} \\
        \cI_C[A] N_S[\rho] &= \cI_{V}[\rho + A] \\
        \cI_V[\rho] N_S[\rho + A] &= \cI_C[A] \\
        \cI_S[A] N_C[\rho] &= \cI_V[\rho + A] (-1)^{\Arf[\rho + A]} \\
        \cI_C[A] N_C[\rho] &= \cI_C[A] (-1)^{\Arf[\rho + A] + \Arf[\rho]} \\
        \cI_V[\rho] N_C[\rho + A] &= \cI_S[A] (-1)^{\Arf[\rho]}
    \end{split}
\end{equation}

\subsubsection*{Invertible Defects and Topological Manipulations}

\begin{figure}
    \centering
    \begin{tikzpicture}
        \shade[top color=dgreen!30, bottom color=dgreen!5] (0,0) rectangle (2,3);
        \shade[top color=blue!30, bottom color=blue!5] (2,0) rectangle (6,3);
        \shade[top color=dgreen!30, bottom color=dgreen!5] (9,0+3) rectangle (11,3+3);
        \shade[top color=blue!30, bottom color=blue!5] (11,0+3) rectangle (13,3+3);
        \draw[line width=1pt] (2,0) -- (2,3);
        \draw[line width=1pt] (4,0) -- (4,3);
        \draw[line width=1pt] (6,0) -- (6,3);
        \draw[line width=1pt] (11,0+3) -- (11,3+3);
        \draw[line width=1pt] (13,0+3) -- (13,3+3);
        \node at (7.5,1.5+2) {\rotatebox[origin=c]{30}{$\stackrel{\text{1st of \eqref{eq:operinterface1}}}=$}};
        \node at (7.5,1.5-2) {\rotatebox[origin=c]{-30}{$=$}};
        \node at (11,1.5) {\rotatebox[origin=c]{90}{$\stackrel{ \eqref{eq:TcTS}}=$}};
        \node[below] at (2,0) {$\cI_S[A]$};
        \node[below] at (4,0) {$N_V$};
        \node[below] at (6,0) {$\ket{\cZ_F}$};
        \node[below] at (11,0+3) {$\cI_C[A]$};
        \node[below] at (13,0+3) {$\ket{\cZ_F}$};
        \node[right] at (9,0.5) {$\frac{1}{\sqrt{|H^1|}} \sum_a (-1)^{\int_\Sigma a \smile A} \times {}$};
        \shade[top color=dgreen!30, bottom color=dgreen!5] (9,0-3) rectangle (11,3-3);
        \shade[top color=blue!30, bottom color=blue!5] (11,0-3) rectangle (13,3-3);
        \draw[line width=1pt] (11,0-3) -- (11,3-3);
        \draw[line width=1pt] (13,0-3) -- (13,3-3);
        \node[below] at (11,0-3) {$\cI_S[A]$};
        \node[below] at (13,0-3) {$\ket{\cZ_F}$};
    \end{tikzpicture}
    \caption{Gauging $\bZ_2$ symmetry is realised by inserting a topological surface operator $N_V$ in the slab of a bosonic theory.}
    \label{fig:KWslab}
\end{figure}
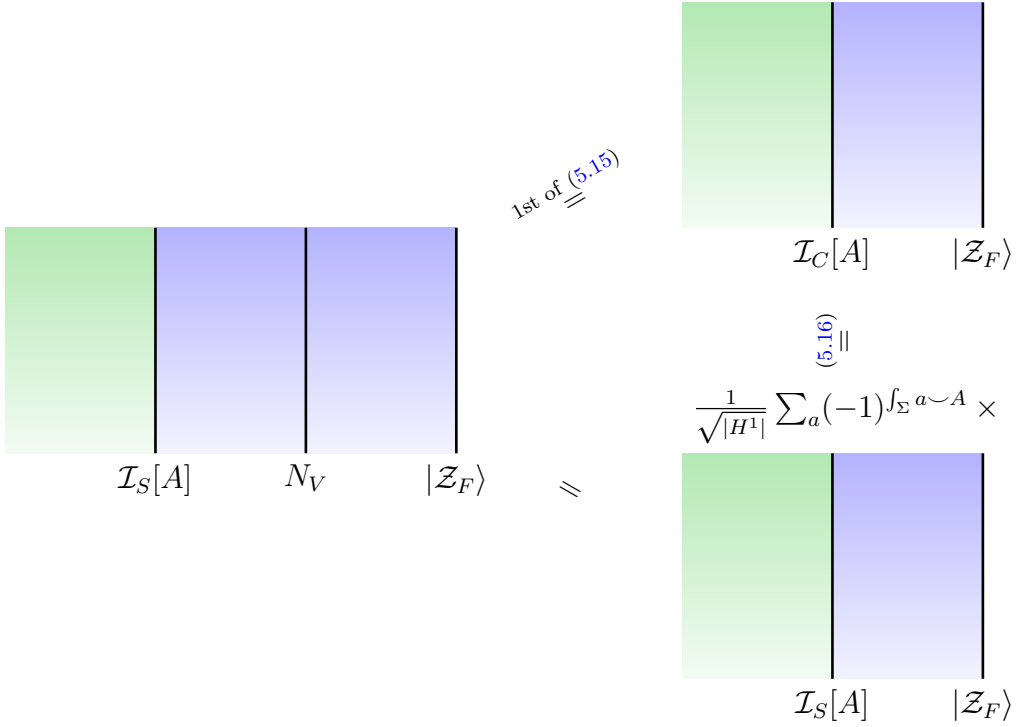

Having explained how the (almost) topological boundary conditions/interfaces are exchanged under fusing with the bulk invertible defects, we finally relate them to the topological manipulations, including bosonisation, fermionisation, gauging $\bZ_2$ symmetry, and stacking an Arf. This can be achieved by taking the inner product with the dynamical boundary state $\ket{\cZ_F}$ on both sides of \eqref{eq:operinterface1}. For instance, denote by $\cT_{S,C}$ the bosonic theory associated with the topological interface $\cI_{S,C}[A]$ respectively. On one hand, by the 1st equality in \eqref{eq:operinterface1}, the two bosonic theories $\cT_S$ and $\cT_C$ are related by inserting a topological surface operator $N_V$ in the slab. On the other hand, using \eqref{eq:interfacerelations}, we have
\begin{equation}\label{eq:TcTS}
    \cZ_{\cT_C}[A] = \frac{1}{\sqrt{|H^1(\Sigma; \bZ_2)|}} \sum_{a \in H^1(\Sigma; \bZ_2)} (-1)^{\int_\Sigma a \smile A} \cZ_{\cT_S}[a]
\end{equation}
The combination of the above two relations shows that inserting $N_V$ in the slab associated with a bosonic theory amounts to a Kramers-Wannier transformation, i.e.\ gauging $\bZ_2$ global symmetry. See Figure \ref{fig:KWslab} for a graphical illustration.

Similarly, by the 3rd equality of \eqref{eq:operinterface1}, the two fermionic theories $\cT_V$ and $\cT_V (-1)^{\Arf[\rho]}$ are also related by inserting a topological surface operator $N_V$ in the slab, so $N_V$ acts on a fermionic theory as stacking an Arf invariant $(-1)^{\Arf[\rho]}$. This reproduces the discussions in Section \ref{sec:propSymTFT}, as shown in Figure \ref{fig:stacking}.

\begin{figure}
    \centering
    \begin{tikzpicture}
        \shade[top color=dgreen!30, bottom color=dgreen!5] (0,0) rectangle (2,3);
        \shade[top color=blue!30, bottom color=blue!5] (2,0) rectangle (6,3);
        \shade[top color=dgreen!30, bottom color=dgreen!5] (9,0+3) rectangle (11,3+3);
        \shade[top color=blue!30, bottom color=blue!5] (11,0+3) rectangle (13,3+3);
        \draw[line width=1pt] (2,0) -- (2,3);
        \draw[line width=1pt] (4,0) -- (4,3);
        \draw[line width=1pt] (6,0) -- (6,3);
        \draw[line width=1pt] (11,0+3) -- (11,3+3);
        \draw[line width=1pt] (13,0+3) -- (13,3+3);
        \node at (7.5,1.5+2) {\rotatebox[origin=c]{30}{$\stackrel{ \eqref{eq:VSS}}=$}};
        \node at (7.5,1.5-2) {\rotatebox[origin=c]{-30}{$=$}};
        \node at (11,1.5) {\rotatebox[origin=c]{90}{$\stackrel{ \eqref{eq:TSTV}}=$}};
        \node[below] at (2,0) {$\cI_V[\rho]$};
        \node[below] at (4,0) {$N_S[\rho+A] N_V$};
        \node[below] at (6,0) {$\ket{\cZ_F}$};
        \node[below] at (11,0+3) {$\cI_S[A]$};
        \node[below] at (13,0+3) {$\ket{\cZ_F}$};
        \node[right] at (9,0.5) {$\frac{1}{\sqrt{|H^1|}} \sum_\rho (-1)^{\Arf[\rho + A] + \Arf[\rho]} \times {}$};
        \shade[top color=dgreen!30, bottom color=dgreen!5] (9,0-3) rectangle (11,3-3);
        \shade[top color=blue!30, bottom color=blue!5] (11,0-3) rectangle (13,3-3);
        \draw[line width=1pt] (11,0-3) -- (11,3-3);
        \draw[line width=1pt] (13,0-3) -- (13,3-3);
        \node[below] at (11,0-3) {$\cI_V[\rho]$};
        \node[below] at (13,0-3) {$\ket{\cZ_F}$};
    \end{tikzpicture}
    \caption{Bosonisation is realised by inserting a topological surface operator $N_S[\rho+A] N_V$ in the slab of a fermionic theory. }
    \label{fig:Bosslab}
\end{figure}
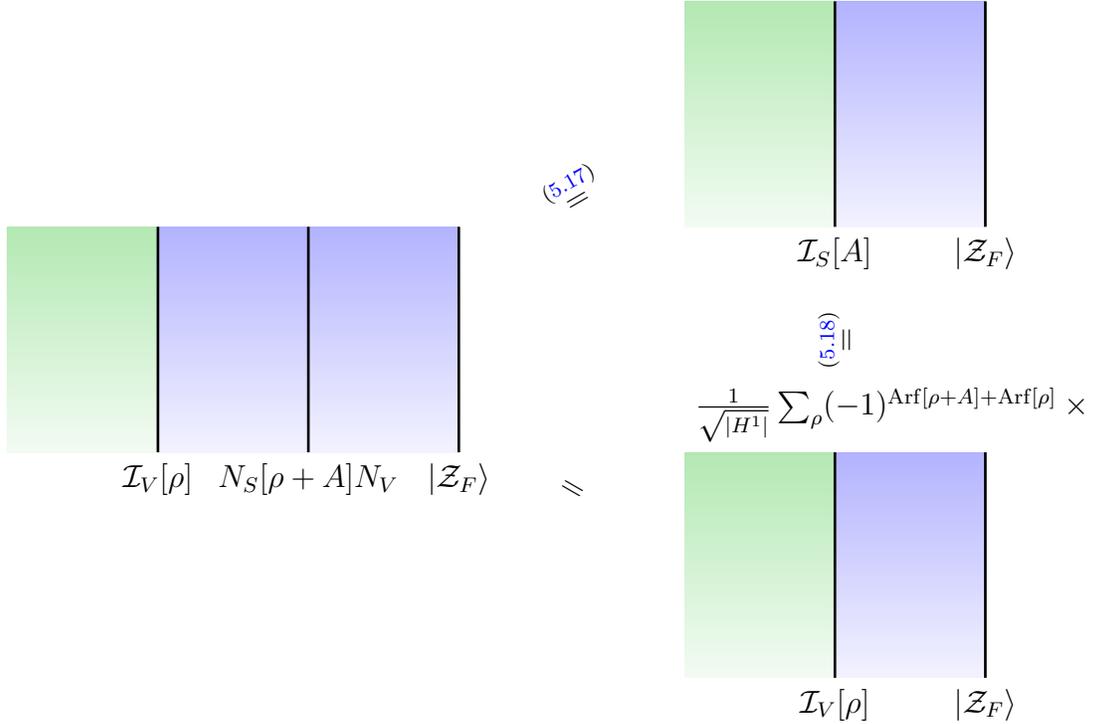

To see how the bosonisation map is realised, we first note that
\begin{equation}\label{eq:VSS}
    \cI_V[\rho] N_S[\rho + A] N_V = \cI_S[A]
\end{equation}
This means the fermionic theory $\cT_V$ and the bosonic theory $\cT_S$ are related by inserting a topological surface operator $N_S[\rho+A] N_V$ in the slab. On the other hand, using the inverse transformation of the second in \eqref{eq:interfacerelations}, we have
\begin{equation}\label{eq:TSTV}
    \cZ_{\cT_S}[A] = \frac{1}{\sqrt{|H^1(\Sigma; \bZ_2)|}} \sum_\rho (-1)^{\Arf[\rho + A] + \Arf[\rho]} \cZ_{\cT_V}[\rho]
\end{equation}
Comparing with \eqref{eq:spinsumdef}, we conclude that inserting $N_S[\rho + A] N_V$ in the slab associated with a fermionic theory realises a bosonisation map. See Figure \ref{fig:Bosslab} for a graphical illustration. Likewise, one can also show that the fermionisation map is achieved by inserting $N_C[\rho] N_V$ into the slab associated with the theory $\cT_S$.

\subsubsection*{Connection to Non-Invertible Condensation Defects}

The invertible defect $N_S[\rho]$ is almost topological. One can make it topological by summing over the spin structure. Using the Arf invariant identities, we find
\begin{equation}
    \sum_\rho N_S[\rho] = \cZ_{D(\bZ_2)} \cN_S
\end{equation}
where $\cZ_{D(\bZ_2)} = \sqrt{|H^1(\Sigma; \bZ_2)|}$ is the partition function of the $\bZ_2$ gauge theory, and $\cN_S$ is a spin structure \emph{independent} topological defect obtained by condensing the bosonic line $L_S$ on $\Sigma$,
\begin{equation}
    \cN_S = \frac{1}{\sqrt{|H^1(\Sigma; \bZ_2)|}} \sum_{\gamma \in H_1(\Sigma; \bZ_2)} L_S(\gamma)
\end{equation}
$\cN_S$ satisfies the non-invertible fusion rule $\cN_S \times \cN_S = \cZ_{D(\bZ_2)} \cN_S$ \cite{Roumpedakis:2022aik}. This illustrates how a defect can be made topological at the expense of sacrificing invertibility, providing an alternative explanation for the non-invertibility of $\cN_S$ motivated by bosonisation.

\subsubsection*{Relation with $\bZ_2$ Gauge Theory}

We conclude this section by commenting on the relation between the interfaces discussed in this subsection, and the boundary conditions of the $\bZ_2$ gauge theory $D(\bZ_2)$. When $n = 16n_B$, the SymTFT obeys a useful duality:
\begin{equation}\label{eq:duality}
    \Spin(16n_B)_{-1} \longleftrightarrow D(\bZ_2) \times \Spin(16n_B)_{-1} / \bZ_2^S
\end{equation}
Here $D(\bZ_2)$ is an (untwisted) $\bZ_2$ gauge theory, and $\Spin(16n_B)_{-1} / \bZ_2^S$ is the bosonic SPT we met in Section~\ref{sec:grav}. This duality cleanly separates the TFT nature, which is captured by $D(\bZ_2)$, from the gravitational anomaly, which is captured by $\Spin(16n_B)_{-1} / \bZ_2^S$. All the interesting physics is in the $D(\bZ_2)$ factor. This is why there is little difference between $n_B \neq 0$ and $n_B = 0$.

The theory $D(\bZ_2)$ has four line operators, usually denoted as $1, e, m, f$, corresponding to our $1, S, C, V$. There are two topological boundary conditions, Dirichlet $\bra{D[A]}$ and Neumann $\bra{N[A]}$, associated with condensation of the $e$ line and $m$ line respectively. We can import these boundary conditions into the Symmetry TFT using \eqref{eq:duality}. Simply pick a choice of vacuum state $\ket{0}$ for $\Spin(16n_B)_{-1} / \bZ_2^S$, and form the tensor product $\bra{D[A]}\bra{0}$. But echoing a by now familiar story, the state $\ket{0}$ is phase-ambiguous due to the gravitational anomaly, and so in order to get an unambiguous answer we should instead form the operator $\ket{0} \bra{D[A]}\bra{0}$.\footnote{This is actually none other than $\bra{D[A]}$ itself, regarded as an operator.} This is what we have called $\cI_S[A]$. Similarly, $\ket{0} \bra{N[A]}\bra{0}$ is the topological interface $\cI_C[A]$.

\subsection{The Case $n = 8$ mod 16}

Here we turn to our primary interest, the duality web for theories with $n = 8$ mod 16, and its encoding by the Symmetry TFT $\Spin(n)_{-1}$. See also \cite[Section 3.1.1]{orbifoldgroupoids} for early discussions.

\subsubsection*{Lines and Condensation Defects}

When $n = 8$ mod 16, the previously bosonic lines $S$ and $C$ become fermionic. We therefore have a grand total of three fermionic lines
\begin{equation}
    L_V(\gamma) \quad L_S(\gamma) \quad L_C(\gamma)
\end{equation}
which are permuted by an $\bS_3$ symmetry. This symmetry can be understood in terms of the outer automorphism group $\text{Out}(\Spin(8)) \cong \bS_3$, via an analogous duality to \eqref{eq:duality}:
\begin{equation}\label{eq:duality2}
    \Spin(8 + 16n_B)_{-1} \longleftrightarrow \Spin(8)_{-1} \times \Spin(16n_B)_{-1}/\bZ_2^S
\end{equation}
The lines satisfy the commutation relations
\begin{equation}\label{eq:comm8}
    L_I(\gamma) L_J(\gamma') = (-1)^{\braket{\gamma, \gamma'}} L_J(\gamma') L_I(\gamma)
    \qquad I \neq J, \quad I,J = V,S,C
\end{equation}
and the quantum torus algebra
\begin{equation}
    L_I(\gamma + \gamma') = (-1)^{\braket{\gamma,\gamma'}} L_I(\gamma) L_I(\gamma')
    \qquad I = V,S,C
\end{equation}
As with any fermionic lines, the associated condensation defects
\begin{equation}
    N_J = \frac{1}{\sqrt{|H^1(\Sigma; \bZ_2)|}} \sum_{\gamma \in H_1(\Sigma; \bZ_2)} L_J(\gamma)
    \qquad J = V,S,C
\end{equation}
are invertible and square to 1, and satisfy
\begin{equation}\label{eq:NNN}
    N_I N_J = N_J N_{IJ}
    \qquad I \neq J, \quad I,J = V,S,C
\end{equation}
where $N_{IJ}$ is the surface operator associated with condensing $L_I L_J$, namely $N_K$ where $L_I L_J = L_K$. The $N_I$ can be understood in terms of the $\bS_3$ as the generators of the three $\bZ_2$ subgroups.

\subsubsection*{Interfaces With Invertible Theories}

Condensing the fermion line of type $I$ on half of spacetime generates an interface between the $\Spin(n)_{-1}$ and $\SO(n)_{-1}$ theories, which we denote as $\cI_I[\rho]$. Explicitly, $\cI_I[\rho] = \ket{\rho} \bra{\rho}_I$ for some topological boundary conditions $\ket{\rho}_I$, with $\ket{\rho}_V = \ket{\rho}$. However only the interfaces $\cI_I[\rho]$ are well-defined, phase-unambiguous objects.

As in \eqref{eq:IVLV}, $\cI_I[\rho]$ is an eigenstate of $L_I(\gamma)$,
\begin{equation}
    \cI_I[\rho] L_I(\gamma) = \cI_I[\rho] (-1)^{\Arf[\rho + \PD(\gamma)] + \Arf[\rho]}
\end{equation}
for $I = V,S,C$. This immediately implies
\begin{equation}\label{eq:IILI}
    \cI_I[\rho] N_I = \cI_I[\rho] (-1)^{\Arf[\rho]}
\end{equation}
How about $\cI_I[\rho] N_J$ for $I \neq J$? We first note
\begin{equation}
    \cI_I[\rho] N_J N_{IJ} = \cI_I[\rho] N_J (-1)^{\Arf[\rho]}
\end{equation}
This implies that $\cI_I[\rho] N_J$ is proportional to $\cI_{IJ}[\rho]$, where $\cI_{IJ}$ is the interface associated with condensing $L_I L_J$. The most general proportionality constant one can write down is $(-1)^{p_J \Arf[\rho]}$. Concretely,
\begin{equation}\label{eq:generans}
    \begin{split}
        \cI_S[\rho] N_C &= \cI_V[\rho] (-1)^{p_C \Arf[\rho]} \\
        \cI_C[\rho] N_V &= \cI_S[\rho] (-1)^{p_V \Arf[\rho]} \\
        \cI_V[\rho] N_S &= \cI_C[\rho] (-1)^{p_S \Arf[\rho]}
    \end{split}
\end{equation}
Since the $V,S,C$ operators are all on equal footing, the three equations should be mapped to each other under the permutation $V \rightarrow S \rightarrow C \rightarrow V$, which implies that the three unknown coefficients reduce to one, $p_S = p_C = p_V = p$. Combining with \eqref{eq:IILI} and \eqref{eq:NNN}, one further derives\footnote{For example, the first equality can be derived as follows: $\cI_S[\rho] N_V \stackrel{\eqref{eq:NNN}}{=} \cI_S[\rho] N_C N_S N_C \stackrel{\eqref{eq:generans}}{=} (-1)^{p \Arf[\rho]} \cI_V[\rho] N_S N_C \stackrel{\eqref{eq:generans}}{=} \cI_C[\rho] N_C \stackrel{\eqref{eq:IILI}}{=} \cI_C[\rho] (-1)^{\Arf[\rho]}$.}
\begin{equation}\label{eq:generansz2}
    \begin{split}
        \cI_S[\rho] N_V &= \cI_C[\rho] (-1)^{\Arf[\rho]} \\
        \cI_C[\rho] N_S &= \cI_V[\rho] (-1)^{\Arf[\rho]} \\
        \cI_V[\rho] N_C &= \cI_S[\rho] (-1)^{\Arf[\rho]}
    \end{split}
\end{equation}
Combining \eqref{eq:generans} and \eqref{eq:generansz2} we find $p=1$. To see this, for instance, right-multiplying by $N_C$ on both sides of the first equality in \eqref{eq:generans}, one finds $\cI_S[\rho] = \cI_V[\rho] N_C (-1)^{p \Arf[\rho]}$. Further applying the third equality in \eqref{eq:generansz2}, the above reduces to $\cI_S[\rho] = \cI_S[\rho] (-1)^{(p+1) \Arf[\rho]}$, which is consistent only when $p=1$. Indeed those in \eqref{eq:generans} and \eqref{eq:generansz2} are related by $\bZ_2 \subset \bS_3$, say $S \leftrightarrow C$ while keeping $V$ fixed.

\subsubsection*{Duality Map}

After capping the dynamical boundary state $\ket{\cZ_F}$ with the interfaces $\cI_{V,S,C}$, one obtains three fermionic theories denoted $\cT_{V,S,C}$. Inserting the invertible condensation defects $N_{V,S,C}$ amounts to performing topological manipulations, which we aim to identify.

Starting with the theory $\cT_I$, inserting $N_I$ is clearly identified with stacking an Arf invariant. However, starting with theory $\cT_V$ for instance, which invertible condensation defect shall we interpret as the refermionisation transformation? First note that such an operation is of order 2 as explicitly demonstrated in \ref{sec:cft}, and should map $\cT_V$ to either $\cT_S$ or $\cT_C$, possibly stacked with an Arf invariant. There are two operators satisfying these conditions: $N_S$ and $N_C$. This is as expected; there are two refermionisation maps $R^{(1)}$ and $R^{(2)}$, as discussed in detail in Section \ref{sec:anomalies}. Hence we obtain the same duality web as in \eqref{eq:dualityweb}, where $\cT_V \leftrightarrow T_{F_1}, \cT_S \leftrightarrow T_{F_2}, \cT_C \leftrightarrow T_{F_3}$. We also see a reflection of the relation \eqref{eq:Rrelation}, which manifests here as the identity $N_S = N_V N_C N_V$.

\begin{figure}
    \centering
    \begin{tikzcd}[column sep=large, row sep=large]
        & \cT_V \times \Arf \arrow[r, blue, leftrightarrow, "N_C"] & \cT_S \arrow[dr, red, leftrightarrow, "N_S"] & \\
        \cT_V \arrow[ur, red, leftrightarrow, "N_V"] & & & T_S \times \Arf \arrow[dl, blue, leftrightarrow, "N_V"] \\
        & \cT_C \times \Arf \arrow[ul, blue, leftrightarrow, "N_S"] & \cT_C \arrow[l, red, leftrightarrow, "N_C"] &
    \end{tikzcd}
    \caption{Duality web for $n = 8$ mod 16, derived from the Symmetry TFT.}
    \label{fig:dualitywebsymtft}
\end{figure}
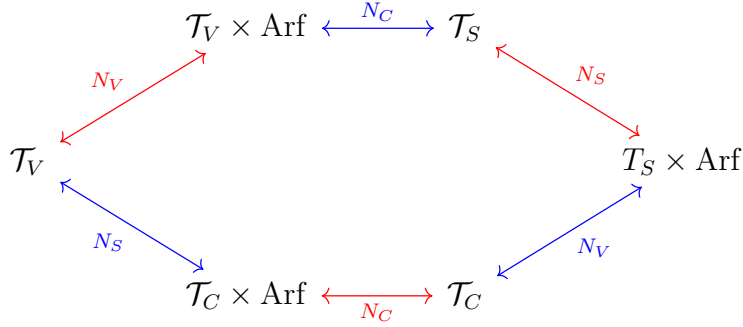

We end this section by commenting that it is possible to generalise the discussion in the current section to 3d/4d systems. As remarked at the end of Section \ref{sec:anomalies}, the spin structure in 3d is anomaly-free, hence one can sum over the spin structure without problem. The spin structure sum can also be captured by changing the topological boundary condition of the SymTFT---(fermionic) $\bZ_2$ gauge theory in 4d. Various spin structure-dependent boundary conditions and topological defects have recently been investigated \cite{Barkeshli:2023bta}, and interesting categorical structures among defects in diverse dimensions have been identified. It would be interesting to investigate how these findings play a role in bosonisation.

\subsubsection*{Matrix Elements}

We have seen that the theories before ($F$) and after ($F'$) gauging fermion parity are related via\footnote{Here $N_S$ represents one of the two possible definitions of gauging fermion parity; the other would be represented by $N_C$.}
\begin{align*}
    \cZ_F[\rho] &= \cI_V[\rho] \ket{\cZ_F} \\
    \cZ_{F'}[\rho] &= \cI_V[\rho] N_S \ket{\cZ_F}
\end{align*}
The reader may wonder why we have not provided explicit matrix elements for $N_S$. After all, this information would come in pretty useful if we wanted to concretely compute $\cZ_{F'}[\rho]$ from $\cZ_F[\rho]$, which we would do via
\[
    \cZ_{F'}[\rho'] = \sum_\rho \ket{\rho'} \braket{\rho' | N_S | \rho} \bra{\rho} \cZ_F[\rho]
\]
The reason is that this is not a well-defined question. Recall that the spaces $\cH_{\SO(n)_{-1}}[\rho]$ for different $\rho$ are only canonically isomorphic up to a sign. This means any question involving the relative phase between two different $\ket{\rho}$ states is sign-ambiguous, and that includes the numerical values of $\braket{\rho' | N_S | \rho}$.

One way to get a concrete answer is to pick reference data. For example, in Section~\ref{sec:cft}, the CFT regularisation implicitly made such a choice for us, and this allowed us to write down the concrete equation \eqref{eq:elusive_matrix_elements} which encodes the matrix elements of $N_S$. We see that each one is equal to $\pm \frac{1}{|H^1(\Sigma; \bZ_2)|}$.

Alternatively, the gauge invariant content of $\braket{\rho' | N_S | \rho}$ at any genus can be extracted using the equation
\begin{equation}\label{eq:coycle}
    \braket{\rho_1 | N_S | \rho_2} \braket{\rho_2 | N_S | \rho_3} \braket{\rho_3 | N_S | \rho_1} = \frac{(-1)^{\Arf[\rho_1 + \rho_2 + \rho_3] + \Arf[\rho_1] + \Arf[\rho_2] + \Arf[\rho_3]}}{|H^1(\Sigma; \bZ_2)|^3}
\end{equation}
Each basis state appears exactly twice in such a way that all phase ambiguities cancel. This is therefore a covariant equation. We have derived this equation using the expression \eqref{eq:elusive_matrix_elements} for the torus, though it is expected to hold for any surface. Note the peculiar quantity $\rho_1 + \rho_2 + \rho_3$; this in fact defines another spin structure via $\rho_1 + (\rho_2 - \rho_3)$, a relation which is symmetrical among $\rho_1$, $\rho_2$, $\rho_3$.

Equation \eqref{eq:coycle}, reminiscent of a cocycle condition, precisely encodes the information that remains in the matrix after the sign ambiguity in the choice of basis. For example, we can take $\rho_3$ to be a reference spin structure, and freely choose the sign of each matrix element involving $\rho_3$ to be positive. \eqref{eq:coycle} then specifies the phase of the matrix elements between all other spin structures.

\section{Applications}\label{sec:applications}

In this section we conclude with some applications to chiral CFTs. We will examine the action of gauging $(-1)^F$ in such theories with gravitational anomaly $n = 8$ mod 16.

A chiral CFT is characterised by the fact it has $\bar{c} = 0$, and hence has a gravitational anomaly of $n = 2c$. All such theories with $c \leq 16$ are products of the following building blocks \cite{Hohn:2023auw, Rayhaun:2023pgc, chiralfermioniccfts}
\begin{equation*}
\def\arraystretch{1.2}
    \arraycolsep=6pt
    \begin{array}{c||c|c|c|c|c|c|c|c|c}
        c & 0 & \frac{1}{2} & \cellcolor[HTML]{eeeeee} 8 & 12 & 14 & 15 & 15\frac{1}{2} & 16 & \cellcolor[HTML]{eeeeee} 16 \\
        \hline
        \text{theory} & \Arf & \psi & \cellcolor[HTML]{eeeeee} (\mathfrak{e}_8)_1 & \mathfrak{so}(24)_1 & (\mathfrak{e}_7)_1^2 & \mathfrak{su}(16)_1 & (\mathfrak{e}_8)_2 & \mathfrak{so}(16)_1^2 & \cellcolor[HTML]{eeeeee} \mathfrak{so}(32)_1
    \end{array}
\end{equation*}
where $\psi$ denotes a single Majorana-Weyl fermion, and $\mathfrak{g}_k$ denotes the unique CFT of the specified central charge with affine $\hat{\mathfrak{g}}_k$ symmetry. Bosonic theories are highlighted in grey. We have also included the $\Arf$ topological theory as a very degenerate CFT; it has the peculiar property that when it occurs in a product with certain other building blocks, namely $\psi$, $(\mathfrak{e}_7)_1^2$, $\mathfrak{su}(16)_1$, $(\mathfrak{e}_8)_2$, it is absorbed.

We are interested in theories with $n = 8$ mod 16, or equivalently $c = 4$ mod 8. Since the above classification only goes up to $c \leq 16$, we will focus on $c = 4$ and $c = 12$. In the remaining sections, we consider how the duality web \eqref{eq:dualityweb} is realised on these CFTs. See also \cite[Remark D.11]{Rayhaun:2023pgc}.

\subsection{$c$ = 4}

When $c = 4$, the only option for building a chiral fermionic CFT is to take 8 copies of $\psi$, which we denote as $8\psi$. All other building blocks have too high a central charge, except for $\Arf$, but $\Arf$ is absorbed by $\psi$. Thus the theory of 8 Majorana-Weyl fermions is the unique chiral fermionic CFT of $c = 4$.

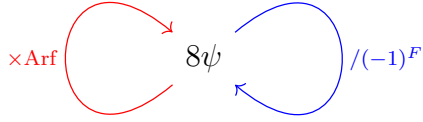
\begin{figure}
    \centering
    \begin{tikzcd}[scale=2]
        8\psi \arrow[red, rightarrow, loop, in=140, out=220, distance=3em, "\times \Arf"] \arrow[blue, rightarrow, loop, distance=3em, in=320, out=40, "/(-1)^F"]
    \end{tikzcd}
    \caption{Web of chiral fermionic CFTs with $c = 4$.}
    \label{fig:orbit1}
\end{figure}

Although the duality web \eqref{eq:dualityweb} in principle contains six different theories, typically some of these theories will coincide, which will lead to the web partially or completely collapsing. In the present case there is only one possible theory, and so the web completely collapses to a single node. See Figure~\ref{fig:orbit1}.

Figure~\ref{fig:orbit1} is of course the statement of the self-triality of 8 Majorana-Weyl fermions. Here we see it emerge as a consequence of the uniqueness of a chiral CFT at $c = 4$, together with the value of its gravitational anomaly, without any mention of $\mathfrak{so}(8)$ triality. On the other hand $\mathfrak{so}(8)$ triality, via \eqref{eq:duality2}, plays a universal role in the duality web of \emph{any} theory with $c = 4$ mod 8, bringing our story full circle.

Note that we view 8 Majorana-Weyl fermions, whether they transform in the $V$, $S$ or $C$ representation of $\Spin(8)$, as a single theory, differing only in how we choose to describe the $\Spin(8)$ global symmetry. This is why Figure~\ref{fig:orbit1} contains only a single node, rather than six. One consequence of this perspective is that each arrow in the diagram involves a reparametrisation of the $\Spin(8)$ symmetry by an outer automorphism. For example, the $\times \Arf$ arrow reparametrises the symmetry by the $S \leftrightarrow C$ automorphism.

The self-duality interface for $/ (-1)^F$ defines a topological interface in the theory $8\psi$ known as the \emph{Maldacena-Ludwig interface} \cite{maldacenaludwig}.\footnote{We thank Masataka Watanabe for pointing this out.} In this sense, the Maldacena-Ludwig interface provides one of the simplest illustrations that gauging $(-1)^F$ can convert a fermionic theory to a fermionic theory.

\subsection{$c$ = 12}

\begin{figure}
    \centering
    \begin{tikzcd}[scale=2, column sep=large, row sep=huge]
        24\psi \arrow[red, rightarrow, loop, in=140, out=220, distance=3em, "\times \Arf"] \arrow[blue, leftrightarrow, r, "/(-1)^F"] &
        \mathfrak{so}(24)_1 \arrow[red, leftrightarrow, r, "\times \Arf"] &
        \mathfrak{so}(24)_1 \times \Arf \arrow[blue, rightarrow, loop, distance=3em, in=320, out=40, shift left=.7em, "/(-1)^F"] \\
        & & (\mathfrak{e}_8)_1 \times 8\psi \arrow[red, rightarrow, loop, in=140, out=220, distance=3em, shift left=.7em, "\times \Arf"] \arrow[blue, rightarrow, loop, distance=3em, in=320, out=40, shift left=.7em, "/(-1)^F"]
    \end{tikzcd}
    \caption{Web of chiral fermionic CFTs with $c = 12$.}
    \label{fig:orbit2}
\end{figure}
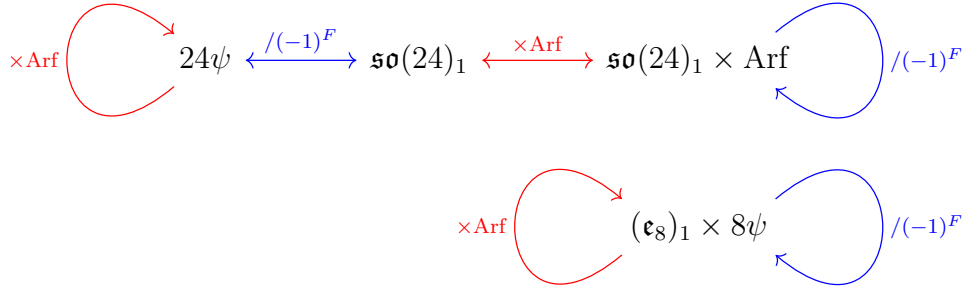

When $c = 12$, there are four possible fermionic chiral CFTs, depicted in Figure~\ref{fig:orbit2}.

The $(\mathfrak{e}_8)_1 \times 8\psi$ theory is the super-version of $(\mathfrak{e}_8)_1$. The bosonic $(\mathfrak{e}_8)_1$ factor plays no role in spin structure manipulations, hence the duality web reduces to that of $8\psi$, and the theory has self-triality.

The $\mathfrak{so}(24)_1$ theory is also known as \emph{Duncan's Supermoonshine module} \cite{duncan2007}. It can be obtained by gauging the $(-1)^F$ of 24 free Majorana-Weyl fermions. Because $24\psi$ is invariant under stacking with $\Arf$, the duality web \eqref{eq:dualityweb} partially collapses to three nodes.

Note that although the diagram appears to single out $\mathfrak{so}(24)_1 \times \Arf$ as being invariant under gauging $(-1)^F$, this is in fact due to the choice of $/ (-1)^F$ as one of the two possible operations $R^{(1)}$, $R^{(2)}$. Making the other choice would exchange the roles of $\mathfrak{so}(24)_1$ and $\mathfrak{so}(24)_1 \times \Arf$, hence their relation is completely democratic.

These four theories have appeared as worldsheet theories of the non-critical two dimensional heterotic string.\footnote{We are grateful to Justin Kaidi for sharing this observation.} In \cite{McGuigan:1991qp, Giveon:2004zz, Davis:2005qe, Seiberg:2005nk}, four heterotic string theories have been identified. They are
\begin{itemize}
    \item HO: The gauge group is $\Spin(24)$ and the spectrum includes massless tachyons in the $\mathbf{24}$ of $\Spin(24)$.
    \item HE: The gauge group is $\Spin(8) \times E_8$ and the spectrum includes 8 massless tachyons in the $\mathbf{8}_v$ representation of $\Spin(8)$.
    \item THO: The gauge group is $\Spin(24)$ and the spectrum does not contain tachyons. There is also an equivalent theory related to THO by spacetime parity.
\end{itemize}
It is known that the number of tachyons in spacetime is the number of free Majorana-Weyl fermions on the worldsheet. See \cite[Section 6]{chiralfermioniccfts} for a recent review. Therefore the HO theory is identified with $24\psi$ on the worldsheet, and the HE theory is identified with $(\mathfrak{e}_8)_1 \times 8\psi$. Among THO and its spacetime parity partner, one of them is identified with $\mathfrak{so}(24)_1$ and the other is identified with $\mathfrak{so}(24)_1 \times \Arf$.

\section*{Acknowledgements}\label{sec:ack}

The authors thank Kantaro Ohmori, Nathan Seiberg and Shu-Heng Shao, who are rumoured to have worked similar results out two years ago, for not publishing. We also thank Yuji Tachikawa and Masataka Watanabe for jointly providing the inspiration for this paper, Jan Albert, Andrea Antinucci, Yichul Choi, Diego Delmastro, Justin Kaidi, Zohar Komargodski, Justin Kulp, Brandon Rayhaun, Shu-Heng Shao, and Adar Sharon for useful discussions, and Jan Albert, Justin Kaidi, Kantaro Ohmori, Yuji Tachikawa, and David Tong for useful comments on the draft. PBS is supported by WPI Initiative, MEXT, Japan at Kavli IPMU, the University of Tokyo.

\appendix

\section{Arf Invariant Identities}\label{app:arf}

We collect some useful formulae obeyed by the Arf invariant. The same formulae can be found, for instance, in \cite{Ji:2019ugf, Karch:2019lnn}. Denote the spin structure on a two dimensional oriented surface $\Sigma$ as $\rho$. We also denote $\bZ_2$ gauge fields by capital letters $A, B, \ldots \in H^1(\Sigma; \bZ_2)$, which can shift the spin structure to $\rho + A$. Then the Arf invariant $\Arf[\rho]$ satisfies the identities
\begin{equation}
    \Arf[\rho + A + B] = \Arf[\rho + A] + \Arf[\rho + B] + \Arf[\rho] + \int_\Sigma A \smile B \quad \text{mod } 2
\end{equation}
and
\begin{equation}
    \frac{1}{\sqrt{|H^1(\Sigma; \bZ_2)|}} \sum_\rho (-1)^{\Arf[\rho]} = 1
\end{equation}

\section{Extended Duality Web}\label{app:web}

Our primary interest has been in fermionic theories with gravitational anomaly $n = 8$ mod 16, and the topological operations one can perform on them. It turns out that similar results hold for fermionic theories with an anomaly-free $\bZ_2$ global symmetry and \emph{arbitrary} gravitational anomaly $n$. Here we comment more on their precise relation.

Given a fermionic theory with an anomaly-free $\bZ_2$ global symmetry, denote gauging the $\bZ_2$ symmetry by $S$, stacking with the SPT $(-1)^{\Arf[\rho + A] + \Arf[\rho]}$ by $T$, and stacking with the SPT $(-1)^{\Arf[\rho]}$ by $Y$. Then these operations satisfy the algebra \cite{bhardwaj2020}
\begin{equation}\label{eq:sty}
    S^2 = T^2 = Y^2 = 1 \qquad (ST)^3 = Y
\end{equation}
Meanwhile, given a fermionic theory with gravitational anomaly $n = 8$ mod 16, the operations one can perform are $R^{(1)}$, $R^{(2)}$, and stacking with the SPT $(-1)^{\Arf[\rho]}$. These are represented by $N_S$, $N_C$ and $N_V$ in the Symmetry TFT, respectively, when inserted between the vector interface $\cI_V[\rho]$ and the dynamical boundary condition $\ket{\cZ_F}$, and obey the $\bS_3$ group algebra.

\begin{figure}
    \centering
    \includegraphics[width=\textwidth]{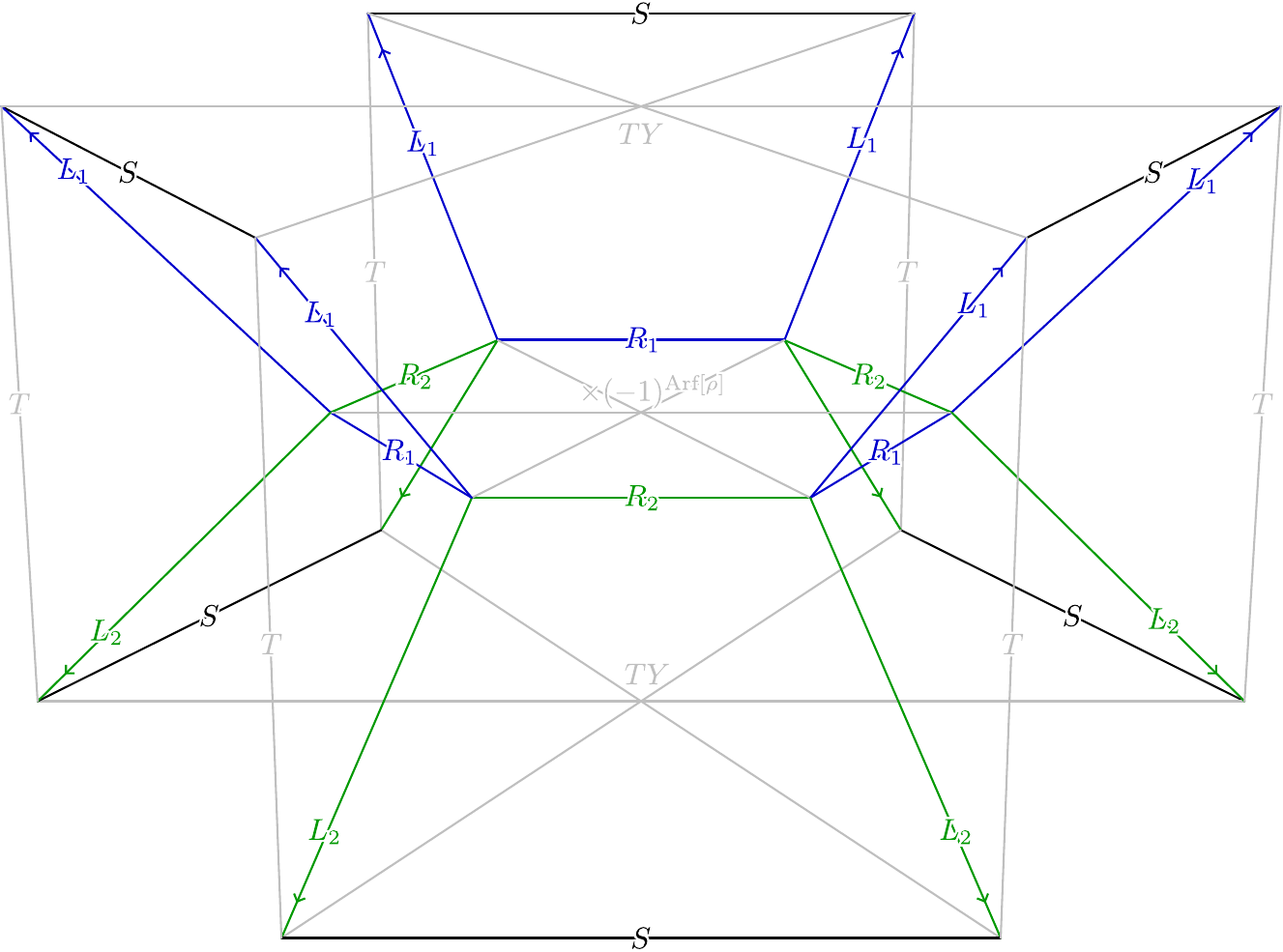}
    \caption{The extended duality web relating the operations studied in this paper to the operations $S$, $T$ and $Y$ via the lifting maps $L_1$ and $L_2$.}
    \label{fig:bigdiagram}
\end{figure}

We would like to relate the two sets of operations above. However, they act on different classes of theories: the first act on partition functions $\cZ_F[\rho, A]$, while the second act on partition functions $\cZ_F[\rho]$. But as discussed in Section~\ref{sec:anomalies}, it is possible to lift $\cZ_F[\rho]$ to $\cZ_F[\rho, A]$ by treating the $(-1)^F$ symmetry as an independent $\bZ_2$ symmetry. Crucially, when the gravitational anomaly is $n = 8$ mod 16, there are two canonically indistinguishable ways to perform the lifting; we shall denote these as $L_1$ and $L_2$. The lifted partition functions are given by
\begin{equation}\label{eq:liftingmatrixelements}
    \cZ_{L_i F}[\rho, A] = (L_i)_{\rho, A} \cZ_{F}[\rho + A]
\end{equation}
for some set of matrix elements $(L_i)_{\rho, A} \in \cH_{\SO(n)_{-1}}[\rho] \otimes \cH_{\SO(n)_{-1}}^*[\rho + A]$. As previously discussed, the two lifts are related by
\begin{equation}
    \begin{tikzcd}[column sep=large, row sep=large]
        & L_1 F \arrow[dd, leftrightarrow, "\times (-1)^{\Arf[\rho + A] + \Arf[\rho]}"] \\
        F \arrow[ur, rightarrow, "L_1"] \arrow[dr, rightarrow, "L_2"] & \\
        & L_2 F
    \end{tikzcd}
\end{equation}
They also obey the obvious identity
\begin{equation}
    \begin{tikzcd}[column sep=15ex, row sep=large]
        L_i F \arrow[r, leftrightarrow, "\times (-1)^{\Arf[\rho + A]}"] & L_i F' \\
        F \arrow[u, rightarrow, "L_i"] \arrow[r, leftrightarrow, "\times (-1)^{\Arf[\rho]}"] & F' \arrow[u, rightarrow, "L_i"]
    \end{tikzcd}
\end{equation}
as a consequence of \eqref{eq:liftingmatrixelements}. Finally, the refermionisation maps $R^{(i)}$ are defined in Section \ref{sec:gauging} as the operation that makes the following square commute:
\begin{equation}
    \begin{tikzcd}[column sep=large, row sep=large]
        L_i F \arrow[r, leftrightarrow, "/ A"] & L_i F' \\
        F \arrow[u, rightarrow, "L_i"] \arrow[r, dashed, leftrightarrow, "R^{(i)}"] & F' \arrow[u, rightarrow, "L_i"]
    \end{tikzcd}
\end{equation}
We are now in a position to relate the topological operations $S, T, Y$ to those studied in this paper. The above commutative diagrams assemble into an extended duality web, which is shown in Figure~\ref{fig:bigdiagram}.

In the middle layer, we have six fermionic quantum field theories with gravitational anomaly $n = 8$ mod 16. These are related to each other under the operations of $R^{(1)}$, $R^{(2)}$ and stacking with $(-1)^{\Arf[\rho]}$, obeying the $\bS_3$ group algebra. Lifting these using $L_1$ results in the six fermionic quantum field theories with $\bZ_2$ symmetry in the top layer, while lifting them using $L_2$ results in the six in the bottom layer. The twelve theories in the top and bottom layers are then exchanged among each other by the operations $S$, $T$ and $Y$ in such a way that the diagram commutes.


\bibliographystyle{jhep}
\bibliography{refs}

\end{document}